\documentclass[12pt,preprint]{aastex}
\newcommand{\hi}{H~{\small{I}}}
\newcommand{\hifig}{H~{\scriptsize{I}}}
\newcommand{\co}{CO(J=1$-$0)}
\newcommand{\tabfrac}[2]{%
	\setlength{\fboxrule}{0pt}%
	\fbox{$\frac{#1}{#2}$}%
}

\shorttitle{Interferometric Maps of \hi\ and CO in LCBGs}
\shortauthors{Garland et al.}

\begin{document}

%% Title
\title{The Nature of Nearby Counterparts to Intermediate-Redshift Luminous 
Compact Blue Galaxies.  III.  Interferometric Observations of Neutral 
Atomic and Molecular Gas}

%% Authors and contact information
\author{C. A. Garland}
\affil{Natural Sciences Department, Black Science Center, Castleton State 
College, Castleton, VT 05735}
\email{catherine.garland@castleton.edu}

\author{D. J. Pisano\altaffilmark{1,2}}
\affil{Remote Sensing Division, Code~7213, Naval Research Lab, 
Washington, DC 20375-5351}
\email{dpisano@nrao.edu}
\altaffiltext{1}{National Research Council Postdoctoral Fellow }
\altaffiltext{2}{Current Address: National Radio Astronomy Observatory, P.O. 
Box 2, Green Bank, WV 24944; dpisano@nrao.edu}

\author{J. P. Williams}
\affil{Institute for Astronomy, University of Hawai\textquoteleft{i}, 
2680~Woodlawn Drive, Honolulu, HI 96822}
\email{jpw@ifa.hawaii.edu}

\author{R. Guzm\'an}
\affil{Department of Astronomy, University of Florida, 211 Bryant 
Space Science Center, P.O.~Box~112055, Gainesville, FL 32611}
\email{guzman@astro.ufl.edu}

\author{F. J. Castander}
\affil{Institut de Ci\`encies de l'Espai (IEEC/CSIC), Campus UAB, 08193 
Bellaterra, Barcelona, Spain}
\email{fjc@ieec.fcr.es}

\and

\author{Leslie J. Sage}
\affil{Department of Astronomy, University of Maryland, College Park, MD 20742}
\email{lsage@astro.umd.edu}

%% Abstract
\begin{abstract}
We present the results of a Very Large Array (VLA) and Owens Valley Radio 
Observatory Millimeter Wavelength Array (OVRO--MMA) follow-up to our
single-dish surveys of the neutral atomic and molecular gas in a
sample of nearby Luminous Compact Blue Galaxies (LCBGs).  These
luminous, blue, high surface brightness, starbursting galaxies were
selected using criteria similar to that used to define LCBGs at higher
redshifts. The surveys were undertaken to study the nature and
evolutionary possibilities of LCBGs, using dynamical masses and gas
depletion time scales as constraints.  Here we present nearly resolved
VLA \hi\ maps of four LCBGs, as well as results from the literature for a
fifth LCBG.  In addition, we present OVRO--MMA maps of CO(J=1$-$0) in two of
these LCBGs.  We have used the resolved \hi\ maps to separate the \hi\
emission from target galaxies and their companions to improve the
accuracy of our gas and dynamical mass estimates. For this sub-sample
of LCBGs, we find that the dynamical masses measured with the
single-dish telescope  and interferometer are in agreement.  However,
we find that we have  overestimated the mass of \hi\ in two galaxies
by a significant amount, possibly as much as 75\%, when compared to
the single-dish estimates. These two galaxies have companions  within
a few arc minutes;  we find that our single-dish and interferometric
measurements of  \hi\ masses are in reasonable agreement for galaxies
with more distant companions. Our  CO(J=1$-$0) maps, despite long
integration times, were faint and barely resolved making analysis
difficult except to verify the central concentration of the molecular
gas.  The \hi\ velocity fields indicate that all five galaxies are
clearly rotating yet distorted, likely due to recent interactions.
Our measurements of the gas and dynamical masses of LCBGs point
towards evolution into low  mass galaxies such as dwarf ellipticals,
irregulars, and low mass spirals, consistent with studies of LCBGs at
higher redshifts.
\end{abstract}

%% Keywords


%% Begin body of paper.

\section{INTRODUCTION}
Luminous Compact Blue Galaxies (LCBGs) are small galaxies brightened by a burst of star formation. Interactions or mergers may have triggered such galaxies to enter the LCBG phase \citep{2001A&A...374..800O, 1991ApJ...381...14L}, and, indeed, many nearby LCBGs have companions and show signatures of interactions \citep{2004ApJ...615..689G}.  LCBGs contribute roughly 45\% of the star formation rate (SFR) density and 20\% of the field galaxy number density at a redshift near one; their contribution has decreased by at least a factor of ten today \citep{2004ApJ...617.1004W, 1997ApJ...489..559G, 1997ApJ...489..543P, 1994AJ....108..437M}.  While it is clear that these galaxies are the fastest evolving galaxy population from a redshift of one to today, the details of this evolution are unclear.

Since luminous, compact, starbursting galaxies appear to represent an important phase in the early history of galaxy formation, Jangren et al. (2007, submitted) developed a classification for such galaxies to permit selection over a wide redshift range (see also \cite{2004ApJ...617.1004W} for a description.)  LCBGs are spectroscopically and morphologically diverse, but are separated from local normal galaxies by their small size, high luminosity, high surface brightness, and blue color.  The specific cut-offs in parameter space that define LCBGs have been selected so that they may be observed out to a redshift of one in deep Hubble Space Telescope images.
Note that while they do have optical diameters of a few kpc, LCBGs have higher luminosities and metalicities than the
Blue Compact Galaxies (BCGs) and Blue Compact Dwarf Galaxies (BCDGs) widely studied by, for example,
\citet{1981ApJ...247..823T}, \citet{1994AJ....107..971T}, and \citet{2002AJ....124..191S}.  Previous BCG and BCDG surveys have used many different selection criteria, leading to various definitions of these galaxies and samples with a range of properties.  Most surveys focused on \emph{dwarf} star forming galaxies by selecting faint sources from emission line surveys.  There are only a few local \textbf{L}CBGs in previous BCG and BCDG surveys. 

Galaxies undergoing an LCBG phase will fade and redden once the star formation activity has been quenched. Their dynamical masses, however, are expected to remain largely unchanged.  Based on measurements of optical sizes and line widths of intermediate redshift LCBGs, it has been suggested that some LCBGs may be the progenitors of today's dwarf ellipticals\footnote{Sometimes called spheroidals.  These are galaxies such as NGC~205, not dwarf spheroidals such as Draco and Carina.} \citep{1996ApJ...460L...5G, 1994ApJ...427L...9K}, irregular, or late type spirals \citep{2006ApJ...640L.143N, 1999ApJ...518L..83M, 1997ApJ...489..543P, 1996ApJ...460L...5G}, and/or spheroidal components of today's disk galaxies \citep{2001ApJ...550..570H, 1997ApJ...489..543P}.  It is unclear, however, if these line widths truly reflect the dynamical masses of the galaxies as such optical emission lines may originate primarily from the central regions of the galaxies.

Measures of the neutral interstellar medium (ISM) in LCBGs provide another avenue to examine the nature and evolutionary possibilities of such galaxies  \citep{2001AJ....122.1194P}.  Neutral atomic hydrogen (\hi) provides a more robust measure of the dynamical mass as it samples the gravitational potential out to a larger scale than optical emission lines.  The mass of atomic and molecular hydrogen, when combined with the SFR, allows an estimate of the length of the starburst activity.  With these points in mind, we began a survey of the neutral ISM in LCBGs.  Current technology limited us to studying the rare, nearby LCBGs.  However, we carefully selected a local sample of LCBGs to match the observed properties (luminosity, surface brightness, and color) of the common, more distant LCBGs so that our findings may apply to both LCBG populations.  

In Paper I \citep{2004ApJ...615..689G}, we described the sample of 20 local (within 70~Mpc) LCBGs, and the results of a single-dish \hi\ survey completed with the Green Bank Telescope (GBT).  Paper II \citep{2005ApJ...624..714G} described a single-dish beam-matched survey of the three lowest rotational transitions of carbon monoxide (CO) in the same sample of galaxies. 
We found LCBGs to be gas-rich, with centrally concentrated CO and low ratios of H$_2$ to \hi\ masses (typically 5\%).  The total gas depletion timescales are short, less than 5 Gyrs for 80\% of our sample. Despite their high luminosities, these LCBGs
have dynamical masses  consistent with low-mass galaxies such as dwarf ellipticals, irregulars, and low luminosity spirals.  These findings are consistent with optical studies of higher redshift LCBGs by, for example, \cite{2006ApJ...640L.143N} and \cite{1996ApJ...460L...5G}.

However, contamination by emission from companion galaxies is a concern, especially for the observations of \hi.  Forty per cent of our sample have companions within the 9$\arcmin$.2 beam of the GBT, potentially leading to overestimates of the gas and dynamical masses.  To improve these measurements, we have begun a program of interferometric, resolved
observations of the \hi\ and CO in our sample of local LCBGs, which we report here.   We assume H$_0$ = 70 km s$^{-1}$ Mpc$^{-1}$ throughout.

\section{OBSERVATIONS AND REDUCTIONS}
To begin our interferometric follow-up, we
selected the brightest local LCBGs with nearby companions in our sample.  We also utilized any available archival and/or published
observations of our original LCBG sample.

\subsection{Neutral Atomic Hydrogen}\label{s:hi_obs}

We used the National Radio Astronomy  Observatory\footnote{The National
Radio Astronomy Observatory is a facility of the  National Science
Foundation  operated under cooperative agreement by Associated
Universities, Inc.}  (NRAO) Very Large Array (VLA) for observations of 21~cm~\hi\ emission.  We observed two
local LCBGs, SDSSJ0834+0139 and Mrk~325.  We obtained data of two
additional sources, Mrk~297 and Mrk~314, from the VLA archives.
Finally, we also used published VLA maps of Mrk~538 from
\cite{1997ApJ...483..754S}.  Observations of the target galaxy and phase
calibrator were alternated, and a flux calibrator observed at the
beginning and/or end of each track.

The observing details for each source, except Mrk~538, are shown in
Table \ref{t:VLAOVRO}, where we list the observing date, array
configuration, flux calibrators, central velocity (heliocentric, except as noted),  total channels, channel
size, and integration time for each source.  The integration time
indicated is the total amount of time.  Typically, the  target was
observed for 30~$-$~40 minutes at a time, the phase calibrator for
four minutes at a time, and the flux calibrator for 10~$-$~30 minutes.
Observations were made during a single day,  except SDSSJ0834+0139,
which was observed on three separate days within a two week period.
The maps of Mrk~538, published by  \cite{1997ApJ...483..754S}, were
made by combining observations in configurations B, C, and D.  The
approximate resolutions of the VLA are 4\arcsec~in B array, 13\arcsec~in C array,
and 10\arcsec~in combined arrays B+C+D.

We used the NRAO 
Astronomical Image Processing System (AIPS) package to edit and calibrate 
the VLA data.  Standard AIPS tasks were used to edit the raw data to 
remove obviously bad sections, calibrate the absolute flux level and 
bandpass shape, and remove continuum emission.  The visibility function 
was then Fast Fourier Transformed (FFT) and the images CLEANed \citep{1974A&AS...15..417H} using the robust weighting algorithm implemented in the AIPS task IMAGR \citep{bri95}.

We found a robustness of zero a good compromise for all sources except
Mrk 314 and Mrk 325, due to their diffuse emission.  
Instead, for Mrk 314, we utilized a robustness of five along with a taper of
15~$\times$~15~k$\lambda$ for imaging.  For Mrk~325, we used a robustness
of two, with no additional taper.  We did not apply tapers to any
other sources.  We examined these initial images to find the channels
containing line emission and created a final set of images using only
those channels.  Finally, we created $\sim$9$^\prime$ $\times$
9$^\prime$ maps of integrated \hi\ intensity (moment 0) and \hi\ velocity (moment 1) using the AIPS task MOMNT.  The maps were blanked at a 3 $\sigma$ level, except for SDSSJ0834+0139 which was not blanked at all due to the low-level emission of its companion galaxy.
Details of the \hi\ intensity maps are shown in Table
\ref{t:VLAOVROmapstats}.  We list the final beam size, the velocity range of the emission, the 1~$\sigma$
noise in the maps, and the corresponding \hi\ mass sensitivities.  Note that diffuse, extended emission was searched
for beyond the $\sim$9$^\prime$ $\times$ 9$^\prime$ regions shown
here, but none was found.  The reduction of the VLA data of Mrk 538 is
described in \cite{1997ApJ...483..754S}; they use pure
natural weighting (robustness = 5).

\subsection{Carbon Monoxide}
The former Owens Valley Radio Observatory Millimeter Wavelength Array
(OVRO--MMA) was utilized to observe Mrk~297 and Mrk~325 in the lowest
rotational transition of carbon monoxide, \co.   The 86~$-$~116 GHz
double-sideband receiver was used with the digital cross-correlator
for these observations.  Observing details, as described in $\S$\ref{s:hi_obs}, are shown in Table \ref{t:VLAOVRO}.  The targets were typically observed for 20 minutes at a time; these observations were alternated with the phase calibrator which was typically observed for four minutes at a time.  In addition, a flux calibrator was observed at the beginning and/or end of each track.
At this frequency the resolution is approximately 10\arcsec~in
C array and 5\arcsec~in 4BG and L array.

We reduced the \co\ data of Mrk~297 and Mrk~325 in the standard manner, similar to the approach described in
\S\ref{s:hi_obs}, but using the mma\footnote{mma is
written and maintained by the California Institute of Technology
millimeter interferometry group specifically for the OVRO--MMA; see \cite{1993PASP..105.1482S}.} data reduction software.  We then used AIPS, as described in
$\S$\ref{s:hi_obs}, to
make maps of the total emission.  Details of the \co\ intensity maps are shown in Table
\ref{t:VLAOVROmapstats}.

\section{RESULTS}
Figures \ref{f:MRK297} through \ref{f:SL0834} show the
total  \hi\ and \co\ intensity maps and intensity-weighted velocity fields of Mrk~297, Mrk~314, Mrk~325, and
SDSSJ0834+0139.  We have restricted the spatial size of the maps to
show only those regions containing significant emission.  We show three \hi\ 
maps for each source: a contour plot of the \hi\ emission down to
3~$\sigma$, an overlay of the R-band Digitized Sky
Survey 2 or r-band Sloan Digital Sky Survey (when available) optical image of the galaxy on a gray scale map of the \hi\ emission, and the \hi\ velocity field.  For Mrk~297 and Mrk~325, we have
also included  maps of the \co\ emission down to a level of 3~$\sigma$,
overlaid on the
\hi\ emission maps.  

\hi\ spectra created from the VLA data using the AIPS task ISPEC are shown in Figure \ref{f:VLAHIspectra}.  Using the same procedures as outlined in Paper~I for our
single-dish spectra, we measured the integrated flux and derived the
corresponding \hi\ mass for each galaxy from these spectra.  Table \ref{t:VLAMHI} lists 
the integrated flux density, $\int$~S~dv, and the corresponding mass
of atomic hydrogen, M$_{HI}$, for each target galaxy.  For comparison, we have also shown the
integrated fluxes and \hi\ masses from the GBT single-dish observations
(Paper~I) and the ratio of the \hi\ masses derived
using the VLA and GBT observations.

Finally, we used the \hi\ spectra to calculate dynamical masses for Mrk~297, Mrk~314, Mrk~325 and SDSSJ0834+0139 using the same procedure detailed in Paper I. (We cannot do detailed velocity field
fits due to the lack of resolution elements across each galaxy.) Table \ref{t:VLAW20} lists inclination and random-motion corrected line widths at 20\% (W$_R^i$), and the derived dynamical masses from the VLA data, assuming virialization.  We also calculated the dynamical mass for Mrk~538, by assuming the spectrum is Gaussian and estimating the
line width at 20\% from the full width at half maximum cited by
\cite{1997ApJ...483..754S}. For comparison, we also list the values derived from the GBT data (Paper I) and the ratio of the VLA and GBT derived dynamical masses. 
Note that the dynamical masses are those within R$_{25}$, the isophotal radius at the limiting surface brightness of 25 B-magnitudes arc sec$^{-2}$.

\subsection{General Results} 
As expected, we find that the \hi\ masses measured using the GBT and VLA are in reasonable agreement for those sources without nearby companions (Mrk~314, Mrk~325) or for mergers (Mrk~297).  Not surprisingly, the single-dish GBT data greatly overestimated the amount of \hi\ in those LCBGs with companions at similar velocities within a few arc minutes (Mrk~538 and SDSSJ0834+0139).

While there were some differences between the line widths measured from the GBT and VLA spectra, in all five cases any differences in the derived dynamical masses are within the random errors, approximately 50\% (Paper~I).  That is, for this small sample of LCBGs, the accuracy of the dynamical mass estimates was not improved by using interferometric data.  The velocity fields of all five galaxies indicate rotation, however the velocity fields are clearly distorted as expected due to the presence of companions.  
As our estimates of dynamical masses assume the galaxies are virialized, we may be overestimating these masses.  For example, in  Mrk~297 we measure a large line width with the VLA which corresponds to a dynamical mass (within R$_{25}$) of 1.7~$\times$~10$^{11}$~M$_\odot$.  However, the velocity field of this galaxy (Figure \ref{f:MRK297.HI.MOM1}) shows a north-south extension (tidal tail) at a high velocity.  This tidal tail contributed at least part of the higher velocity ``shoulder" in the \hi\ spectrum (Figure \ref{f:VLAHIspectra}) used to estimate the galaxy's dynamical mass. 
This resolved interferometric data suggests we are likely overestimating the
rotational velocities of these galaxies.

Despite long integration times, our CO maps of Mrk~297 and Mrk~325, are faint and barely resolved, making further analysis difficult.  However, we did measure the CO(J=1$-$0) radii (R$_{CO}$) for these two galaxies, for comparison with local galaxies of various types.  We chose to measure R$_{CO}$ at a level of
1~K~km~s$^{-1}$, following \cite{1995ApJS...98..219Y}.   This
corresponds to an H$_2$ column density of
1.8~$\times$~10$^{20}$~molecules~cm$^{-2}$, or
$\sim$1~M$_\odot$~pc$^{-2}$ \citep{2001ApJ...547..792D}.  We find  R$_{CO}$ to be
7\arcsec.5~$\times$~4\arcsec.8  in Mrk~297, and
15\arcsec~$\times$~7\arcsec.5  in Mrk~325.  That is,
R$_{CO}$~=~0.30~R$_{25}$ in Mrk~297, and 0.37~R$_{25}$ in Mrk~325.
\cite{1995ApJS...98..219Y} found the ratio of R$_{CO}$ to R$_{25}$ to
be 0.52 $\pm$ 0.02 in a sample of nearby galaxies of varied Hubble
types and environments.  The CO in these two local LCBGs is more
concentrated than the median value found by
\cite{1995ApJS...98..219Y}.  This could be a result of these galaxies undergoing interactions with
companions, which can cause gas to be funneled toward the centers of the primary galaxies \citep[e.g.][]{1996ApJ...460..121W, 1995ApJ...448...41H}.  See Paper II for a complete discussion of the molecular gas in this sample of LCBGs.

\subsection{Individual Galaxies}
\subsubsection{Mrk~297}
Figures \ref{f:MRK297.HI.MOM0} and \ref{f:MRK297.HI.SDSS} show a disturbed,
asymmetric distribution of \hi\ in Mrk~297, with a north-south
elongation.   A slight extension is also visible in the optical image.  The \hi\ velocity field of Mrk~297 (Figure \ref{f:MRK297.HI.MOM1}) shows a
velocity gradient across the main part of the galaxy, but the
north-south extension is at a nearly constant higher velocity.  The measured \hi\ mass, 6.3~$\times$10$^9$~M$_\odot$, is 90\% of that measured with the GBT. We did not expect any contamination as Mrk~297 has no known companions.

Figure
\ref{f:MRK297.HIandCO} shows the \co\ emission originating from a
region between the two peaks of \hi\ emission.  The \co\ emission is compact and offset approximately 5\arcsec~northeast from the optical center.  Our CO map is qualitatively similar to that produced by \cite{1993A&A...273....6S} using the Institut de Radioastronomie Millim\'etrique (IRAM) single-dish telescope, although the \cite{1993A&A...273....6S} map shows more extended CO emission. 

Due to its clumpy, irregular shape, Mrk~297 has been classified as an archetypal Clumpy Irregular Galaxy \citep[e.g.][]{1987IAUS..115..599H}, and was included in the \emph{Atlas of Peculiar Galaxies} \citep{1966apga.book.....A}.
Color maps indicate  star formation is occurring in a number of areas
across most of the galaxy,  while the H$\alpha$ emission is clumpy and
suggestive of  outflows \citep{2001ApJS..136..393C}.  This galaxy has
been interpreted as the result of a collision between two late-type
spirals \citep{1979A&A....78L...5A}, or a spiral and an irregular
\citep{1988Afz....28...47B}, which induced star formation throughout
the galaxy. \cite{1991AJ....101.1601T} found, using numerical N-body simulations, that the optical morphological properties of Mrk~297 can be explained as the result of a co-planar radial penetration collision between two disk galaxies.  They suggest that in this collision, the target galaxy (on left) was deformed into the north-south ``wing'' by the face-on intruder galaxy.  Our \hi\ map (Figure \ref{f:MRK297.HI.SDSS}) shows a tidal tail that is much more extended and lopsided than the optical ``wing'' to which \cite{1991AJ....101.1601T} matched their simulation.  Our \hi\ observations do not appear to support the co-planar radial penetration model; more likely Mrk~297 is in a later stage of merging than suggested by this model.  While our CO maps show molecular gas in the central region of Mrk~297, it is unclear if it is only related to one component of the galaxy as suggested by
\cite{1993A&A...273....6S} and \cite{1990PASJ...42L..45S}.  
 
In our sample of nearby LCBGs, Mrk~297 has
the highest infrared luminosity, dynamical mass, molecular gas mass,
and ratio of molecular to atomic gas mass (Papers I and II).  These
observations may be explained by its suggested merger state:  such a close
interaction could trigger quick conversion of atomic to molecular gas
resulting in a bright starburst, centrally concentrated CO, and a disturbed \hi\ component, as in
Ultraluminous Infrared Galaxies \citep{1988ApJ...334..613S}. 

\subsubsection{Mrk 314}\label{s:Mrk314}
Figure \ref{f:MRK314} shows diffuse,
irregularly shaped \hi\ emission in Mrk~314.  The \hi\ and optical peaks are roughly coincident.  The \hi\ emission
extends well beyond the optical portion of the galaxy, although there
are optical features at the ends of two \hi\ extensions to the south. There is an east-west velocity
gradient across the southern half of the galaxy, while the northern
portion of the galaxy has a confused velocity field (Figure  \ref{f:MRK314.HI.MOM1}). \cite{2002AJ....124..788Y} attempted to map Mrk~314 in CO(J=1--0) using the OVRO--MMA but did not detect the galaxy.  However, their upper limit on a mass of molecular hydrogen was an order of magnitude greater than the mass we detected in our single-dish CO survey (Paper II).    

Mrk~314 is not cataloged as having a companion, but a map of the \hi\ emission made with the VLA in D array by
\cite{1993AJ....105..128T} shows the presence of a companion
5$^\prime$ to the north.  \citet{1993AJ....105..128T} measure an \hi\ mass of 2.5$\times$10$^8$~M$_\odot$ for this companion.  As this low-mass, distant companion is outside the half-power point of the GBT beam, it is no surprise that our measurements of the \hi\ mass using the VLA and GBT are in reasonable agreement:  the VLA detected an \hi\ mass of 2.0~$\times$~10$^9$~M$_\odot$, 80\% of the GBT measured mass.   

Mrk~314 is classified as an elliptical in both the NASA/IPAC Extragalactic
Database (NED) and HyperLeda \citep{2003A&A...412...45P}, and may be a polar ring galaxy, as suggested by
\cite{1990AJ....100.1489W}.  Polar ring galaxies are composed of a gas-rich ring around an early-type, often S0, galaxy, and are most likely
formed by tidal accretion of gas from a companion
\citep{2003A&A...401..817B}.  \cite{2001ApJS..133..321C} obtained deep optical images of Mrk~314 and suggested that optical clumps seen to the south and southwest (lower right) may be part of the polar ring structure.  This is consistent with the \hi\ emission we detect.  We also detect clumps of \hi\ to the east (left) of the main galaxy which are not associated with any optical emission.  It is possible these are also part of the polar ring structure.  If this galaxy is a polar ring galaxy, this may explain this galaxy's unusually
high ratio of gas to dynamical mass (Paper I), the disturbed
appearance of the \hi, and the nearby companion.   Mrk~314 has
the smallest fraction of molecular to atomic hydrogen, and one of the
longest total gas depletion time scales (Paper II) in our sample.  If the \hi\ is not contained within the central region of the LCBG, then it may also imply that Mrk~314 has already consumed its \hi.  Our current observations do not have the spatial resolution to confirm this, however.

\subsubsection{Mrk 325}
Mrk~325 has clumpy star forming regions throughout its disk
\citep{1999ApJ...522..199H, 1976A&A....47..371C}, and is commonly cited as a Clumpy Irregular Galaxy \citep[e.g.][]{1987IAUS..115..599H}, while NED and HyperLeda classify it as a spiral. The disturbed appearance of the \hi\ in Mrk~325, as well as its starburst activity, may be explained by its companion, Mrk~326, roughly 7$^\prime$ to the southeast \citep{2000AJ....119...79C, 1999ApJ...522..199H}.

Figures \ref{f:MRK325.HI.MOM0} and \ref{f:MRK325.HI.DSS}
show centrally peaked, but disturbed, \hi\ emission from Mrk~325. Figure \ref{f:MRK325.HI.MOM1} shows a fairly constant \hi\ velocity, with a small gradient, across
Mrk~325. Figure \ref{f:MRK325.HI.MOM0} also shows the diffuse \hi\ emission of the companion, Mrk~326.  However, note that the \hi\ emission shown for Mrk~326 is not representative of the actual gas distribution as some emission has been subtracted off as part of the continuum subtraction for the target, Mrk~325.  The actual \hi\ distribution of Mrk~326 is symmetric and centered on the optical emission.  

We measured an \hi\ mass of 5.7~$\times$~10$^9$~M$_\odot$ for Mrk~325, and 1.4~$\times$~10$^9$~M$_\odot$ for its companion, or a total system mass of 7.1~$\times$~10$^9$~M$_\odot$.  By comparison, the GBT measured mass of Mrk~325 was 6.3~$\times$~10$^9$~M$_\odot$.  Note that when we correct for the primary beam response of the VLA and the GBT, the total system \hi\ masses agree within the observational errors.  \cite{1997AJ....114...77N} acquired VLA D array observations of Mrk~325 and its companion.  Their observations show more extended emission; they estimate \hi\ masses of 3.6~$\times$~10$^9$~M$_\odot$ for Mrk~325 and 5.0~$\times$~10$^9$~M$_\odot$ for its companion.

The \co\ emission in this galaxy is centrally concentrated and elongated towards the southeast.  Figure \ref{f:MRK325.HIandCO}
shows the \co\ and \hi\ peaks are roughly coincident; the \co\ peak is 
$\sim$8\arcsec~south of the \hi\ peak and slightly offset ($<$ 
3\arcsec~southwest) from the optical position.  This is in contrast to Mrk~297, where the \co\ is between the two \hi\ peaks.

\subsubsection{SDSSJ0834+0139}
Figures \ref{f:SL0834.HI.MOM0} and \ref{f:SL0834.HI.SDSS}
show asymmetric, diffuse \hi\ emission from
SDSSJ0834+0139, and extended, more diffuse \hi\ emission from its companion 
to the
northeast (upper left).  Together, the galaxies are known as
UGC~04480, and are classified as a spiral by NED and Hyperleda, while \cite{1973UGC...C...0000N} described them as a starburst galaxy (SDSSJ0834+0139)
with a lens (the companion) attached to an arm .
The \hi\ from the companion galaxy is much more extended than the
optical emission.  The velocity field (Figure
\ref{f:SL0834.HI.MOM1}) shows an east-west velocity gradient across the
galaxy, while its companion has a constant, higher velocity.  It is
possible that the interaction with SDSSJ0834+0139 has disturbed any
ordered motion that this companion may have had in the past, or that the companion is seen face-on, as suggested by the optical image.

SDSSJ0834+0139 and its companion are less than an arc minute apart, so we expect contamination in our GBT spectrum and corresponding single-dish \hi\ measurements.
Our VLA observations yielded \hi\ masses of 1.4~$\times$~10$^9$~M$_\odot$ for SDSSJ0834+0139, and 2.2~$\times$~10$^9$~M$_\odot$ for its companion.  We show spectra of the target and
its companion in  Figure \ref{f:SL0834decomposed}.  The solid line
shows the broad, low-level emission from SDSSJ0834+0139, the dashed
line the sharply peaked emission from the companion galaxy, and the
dotted line the GBT spectrum, which is the combination of both
sources.  Our GBT data of this system yielded an \hi\ mass of
5.9~$\times$~10$^9$~M$_\odot$, 1.6 times the VLA measurements for the
target and companion combined.  
However, note that the low-level emission from SDSSJ0834+0139 extends
nearly to the edge of the VLA bandpass, $\sim$3980~km~s$^{-1}$, so that we lack a clean continuum channel.  This may result in subtracting some \hi\ emission when we do continuum subtraction.

\subsubsection{Mrk 538}
\cite{1997ApJ...483..754S} show maps of \hi\ emission and the velocity field of
Mrk~538 and its companion.  The \hi\ emission is 
extended, forming a bridge between the two galaxies.
While the inner portion of Mrk~538 shows a normal velocity gradient for a rotating disk, deviations from circular motion are present in the outer portions of the galaxy disk.  The companion galaxy also exhibits signs of rotation \citep{1997ApJ...483..754S}.
The total \hi\ mass for the Mrk~538 system, within a region approximately
4$^\prime$~$\times$~8$^\prime$, is
M$_{HI}$~=~7.0~$\times$~10$^9$~M$_\odot$.  This includes Mrk~538, its
companion, the bridge connecting the two, and other clouds and loops
of \hi\ in the area \citep{1997ApJ...483..754S}.  This total \hi\ mass compares  well with the mass
derived using the GBT, 7.6~$\times$~10$^9$~M$_\odot$.  Due to the small separation between Mrk~538 and its companion, it is not surprising that the mass \cite{1997ApJ...483..754S} report for Mrk~538 alone, 1.7~$\times$~10$^9$~M$_\odot$, is only 22\% of that we reported from the GBT observations.

This pair of galaxies, known as Arp~284, is one of the prototypical 
collisional starburst systems.  
Numerical models and the observed stellar and gas morphologies
suggest a recent collision, between 100 to 200 Myr ago
\citep{2003ApJ...589..157S}.
Multi-wavelength data, summarized in \cite{1997ApJ...483..754S}, reveal 
young, intermediate age, and old stellar populations in Mrk 538. \cite{2001ApJ...552..150L} found, using population synthesis models, 
that the starburst is responsible for only a small
portion of an extended star formation episode, triggered
approximately 10$^8$ years ago. 

\section{DISCUSSION AND CONCLUSIONS}
It is becoming clear, from studies over a range of wavelengths and redshift epochs, that LCBGs are small galaxies undergoing vigorous
starbursts.  The growing consensus on the nature of LCBGs narrows the probable evolutionary pathways these galaxies will follow once they cease their starburst activity.  As LCBGs are heterogeneous, it is likely that multiple evolutionary scenarios apply, including fading and reddening into 
dwarf ellipticals \citep{1996ApJ...460L...5G, 1994ApJ...427L...9K}, evolving into irregulars or late type spirals \citep{2006ApJ...640L.143N, 1999ApJ...518L..83M, 1997ApJ...489..543P, 1996ApJ...460L...5G}, or spheroidal components of today's disk galaxies \citep{2001ApJ...550..570H, 1997ApJ...489..543P}. 

In Papers I and II, we found that the dynamical masses and sizes of these galaxies are consistent with low-mass galaxies such as irregulars, low luminosity spirals, and dwarf ellipticals.  While our interferometric follow-up of a handful of LCBGs indicates that we may be overestimating the dynamical masses of these galaxies, this only strengthens our original conclusions.  As the dynamical masses we measure for LCBGs are lower than typical grand-design spirals, we do not find evidence that LCBGs are spheroidal components of large disk galaxies.

Our entire sample of 20 LCBGs have \hi\ masses ranging from 0.47~$\times$~10$^8$ to
7.9~$\times$~10$^9$~M$_\odot$ (Paper I), within the range of local irregulars and low luminosity spiral galaxies \citep{2000A&A...354..874P, 1994ARA&A..32..115R}.  Local LCBGs also have small (typically less than 5\%) ratios of molecular to atomic hydrogen, consistent with low luminosity spirals \citep{1989ApJ...347L..55Y}.  Our interferometric follow-up has also indicated that we are overestimating the neutral gas mass, and therefore the total gas depletion time scales, of at least some LCBGs.  Again, this strengthens our initial findings:  LCBGs will consume their star formation fuel relatively quickly, typically in less than five Gyrs.  If they consume/expel much of their interstellar medium and halt their star formation, their properties will be consistent with dwarf ellipticals which have low masses of atomic gas ($\lesssim$~10$^6$~M$_\odot$) \citep[e.g.][]{1998ApJ...499..209W, 1997ApJ...476..127Y}.  (However, note that dwarf ellipticals may have higher \hi\ masses than currently assumed;  cluster dwarf ellipticals have been observed with \hi\ masses up to 10$^9$~M$_\odot$ \citep{2006MNRAS.tmp..773M, 2003ApJ...591..167C}.)  

It is unclear if LCBGs will indeed consume and/or expel most of their gas, or if the burst of star formation will halt at some point, leaving a large amount of neutral gas in the galaxy.  Modeling by \cite{1999ApJ...513..142M} and \cite{1994ApJ...431..598D} suggest that starbursts could not expel more than a few percent of the total interstellar medium in galaxies of this size.  Similar arguments led \cite{2004ApJ...617.1004W}, for example, to conclude that LCBGs are not capable of blowing out enough gas to become ``gas-poor'' galaxies, such as dwarf ellipticals, given their typical dynamical masses.  Instead, they suggest LCBGs may fade somewhat but still exhibit modest levels of star formation.  If LCBGs retain much of their interstellar medium and maintain a lower level of star formation, their properties will be consistent with irregulars and low luminosity spirals.

To illustrate these conclusions, in Figure \ref{f:galaxy_comp} we plot the \hi\ mass versus the dynamical mass (within R$_{25}$) of our sample of local LCBGs, as well as a selection of other well known galaxies.  For the five galaxies with VLA observations reported in this paper, we plot both the VLA and GBT derived masses.  For the remainder of the galaxies, we plot simply the GBT derived masses from Paper I. Note that the two galaxies in our sample with the smallest \hi\ masses have uncertain dynamical masses as they are estimated from marginal detections (Paper~I).  The two dashed lines in Figure \ref{f:galaxy_comp} indicate ratios of \hi\ to dynamical mass of 10 and 1\%. For comparison with local LCBGs, we have included well known galaxies of various morphological types:

\begin{itemize}
\item M~31 (Andromeda), a bright, multiple armed spiral member of our Local Group.  It is classified as an Sb in HyperLeda and NED.  We used the \hi\ mass and rotational velocity published by \cite{2006ApJ...641L.109C} when including M~31 in Figure~\ref{f:galaxy_comp}.

\item NGC~55, the brightest member of the Sculptor Group, is classified as a Magellanic barred spiral in HyperLeda and NED.  We used the \hi\ mass and line width published by \cite{1991AJ....101..447P} when including NGC~55 in Figure~\ref{f:galaxy_comp}.

\item NGC~4449 is a global starburst known for having an extended \hi\ distribution, with streamers and filaments extending from the optical disk \citep{1998ApJ...495L..47H}.  It is classified as a barred irregular (HyperLeda) or Magellanic barred irregular (NED). In Figure~\ref{f:galaxy_comp}, we used the \hi\ mass and line width published by \cite{1994A&A...285..385B}.

\item NGC~4640 (VCC 1949) is a member of the Virgo Cluster.  It is classified as an S0 in HyperLeda and a dwarf S0 in NED.  It is also commonly classified as a dwarf elliptical, as, for example, in the Virgo Cluster Catalog \citep{1987AJ.....94..251B}.  We used the \hi\ information culled from the literature by \cite{2003ApJ...591..167C} when including NGC~4640 in Figure~\ref{f:galaxy_comp}. 

\item The Large Magellanic Cloud (LMC), a satellite of the Milky Way with weak spiral features \citep{1955AJ.....60..126D}, is a prototypical Magellanic barred spiral and is classified as such in HyperLeda and NED.  When including the LMC in Figure~\ref{f:galaxy_comp}, we used the \hi\ mass and line width reported by \cite{2003MNRAS.339...87S}.

\item NGC~205 (M~110), a companion to M 31 and classified as elliptical by both HyperLeda and NED. It is also commonly classified as a dwarf Elliptical, e.g., by \cite{1994cag..book.....S}.  We used the \hi\ mass and line width in \cite{1997ApJ...476..127Y} when including NGC~205 in Figure~\ref{f:galaxy_comp}.
\end{itemize}

Dynamical masses for these comparison galaxies were calculated within R$_{25}$ (from HyperLeda, except from \cite{1988AJ.....96..877B} for the LMC) using \hi\ line widths corrected for inclination and random motions in the same manner as described for our LCBG sample in Paper I.  The only exception is M~31, for which we used a rotational velocity derived from an \hi\ rotation curve.  If needed, values were scaled to reflect our use of H$_0$ = 70 km s$^{-1}$ Mpc$^{-1}$.  (More detailed comparisons of the gas and dynamical mass properties with the range of galaxies along the Hubble sequence are explored in Paper I.) 

We have included an example of a non-cluster dwarf elliptical (NGC~205) and a cluster dwarf elliptical (NGC~4640) to illustrate the range in dynamical and \hi\ masses possible for this type of galaxy.  In cluster dwarf ellipticals, the galaxy characteristics are influenced by both internal and external processes, as discussed in, for example, \cite{2005AJ....130.2058B}.  The range of dynamical masses in dwarf ellipticals is still unclear: much depends on environment, but also on differences in data analysis, as pointed out by \cite{2006MNRAS.369.1321D}.  For example, we estimate a dynamical mass within R$_{25}$ of 1.1$\times$10$^8$~M$_\odot$ for NGC~205 by using an \hi\ line width, corrected for turbulence and inclination.  When \cite{2006MNRAS.369.1321D} used models fit to infrared surface brightness distributions and kinematics, a dynamical mass of 1.0$\times$10$^9$~M$_\odot$ within \emph{half} R$_{25}$ was estimated.  Note that \cite{1997ApJ...476..127Y} found that the gas and stars in NGC~205 are kinematically distinct.

For those five galaxies with both GBT and VLA observations, Figure \ref{f:galaxy_comp} illustrates how our interferometric observations have decreased the \hi\ mass estimates, in some cases by a large amount, but have not changed the dynamical mass estimates a great deal.  Looking at Figure~\ref{f:galaxy_comp} in terms of the evolutionary possibilities for LCBGs, it is possible that LCBGs will evolve downward in this figure as \hi\ is transformed into stars (or ejected).  However, it is unlikely, in the framework of passive evolution, that LCBGs will move to the left or right in this figure, since the dynamical mass is assumed to stay relatively constant. From this viewpoint, the likely end-products of LCBGs are the Magellanic spirals and irregulars.  Dwarf ellipticals with high dynamical masses are also a possibility.

Note that while both NGC~55 and NGC~4449 occupy the same region of \hi\ and dynamical mass space in Figure \ref{f:galaxy_comp} as many LCBGs, both are approximately a magnitude too faint (HyperLeda, NED) to be considered \emph{luminous} BCGs. All aspects of LCBGs need to be examined when attempting to trace their evolutionary pathways.  The optical properties, such as stellar masses and color, are also an important consideration and will be discussed by 
Hoyos et al. (2007, submitted.)

The evolutionary scenarios discussed in this paper are consistent with the recent results of 
\cite{2006ApJ...640L.143N} 
who concluded that $\gtrsim$90\% of high redshift (0.2~$\lesssim$~z~$\lesssim$~1.3) LCBGs are small galaxies that will evolve into dwarf ellipticals, irregulars, and small disk galaxies.
However, we note that all these evolutionary scenarios exist in 
the framework of passive evolution; interactions and mergers after the initially triggered burst of star formation may play a 
role in the evolution of at least some LCBGs.  Previous studies have shown 
that major, and potentially even minor, interactions can funnel gas into 
the centers of galaxies and trigger nuclear activity \citep[e.g.][]{1996ApJ...460..121W, 1995ApJ...448...41H, 1984AJ.....89..966D}. Our observations of centrally concentrated CO in our LCBGs are consistent with this scenario.  The asymmetries in the \hi\ profiles and resolved \hi\ maps of 
our LCBGs may be due to ongoing interactions \citep[e.g.][]{1997ApJ...477..118Z, 1994AJ....107.1320O}, but may also be a normal feature of galaxies arising as 
long-lived phenomenon \citep[e.g.][]{2004AJ....127.1900W, 1998ApJ...496L..13L, 1994A&A...290L...9R}.  While interactions and mergers can certainly provide additional 
fuel for star formation and trigger activity in LCBGs, we cannot determine 
the specific effect on our small sample of LCBGs.  However, we can ask, if interactions and mergers provide additional star formation fuel causing LCBGs to keep their ``LCBG'' color, luminosity, and surface brightness characteristics at a later stage in their lives, where are these galaxies today?  In our local universe, LCBGs are quite rare.

We are pursuing interferometric observations of the remaining 15 local LCBGs in our sample using both the VLA and the Giant
Metrewave Radio Telescope (GMRT, India).  The five
LCBGs studied in detail in this paper are among the most luminous of
our sample, and all have companions or are mergers.   Their \hi\ distributions appear
to be quite disturbed, indicating recent or ongoing interactions.
They may not be representative of LCBGs in general, as roughly one
half of our local sample of LCBGs do not show evidence of optical
companions.  The OVRO--MMA maps we presented 
are of two of our brightest galaxies, and yielded barely resolved, faint 
CO emission despite long integration times.  New millimeter telescopes,
such as the GBT, Large Millimeter Telescope (LMT), Combined Array for Research in Millimeter-wave Astronomy (CARMA)
and, eventually, the Atacama Large Millimeter Array (ALMA), will permit more 
detailed studies of the CO in a larger
sample of LCBGs.  Finally, we are also working to compare the detailed \hi\
kinematics with 3D optical spectroscopy for our sample of LCBGs 
\citep{2005sdlb.procP..59P,2006RevMexAA}.  This combined multi-wavelength 
study of 
local LCBGs will provide a key reference to better understand the current
properties and evolutionary pathways of their high-redshift analogs. 

\acknowledgments
We thank the referee for helpful comments, which improved the quality of this paper.  We thank Nicolas Gruel and Jorge P. Gallego for fruitful discussions on 
the evolution of LCBGs.  We also thank Amy McCarty for her assistance comparing the GBT 
and VLA fluxes.  Finally, we thank the staff 
at the NRAO--VLA Array Operations Center and the OVRO--MMA for their help 
with the observations and data reduction.
C. A. G. thanks NRAO for the travel and lodging support while at the 
NRAO--VLA Array Operations Center.  R. G. acknowledges funding from NASA grant LTSA NAG5--11635.  This research was performed in part
while D. J. P. held a National Research Council Research Associateship Award 
at the Naval Research Laboratory.  Basic research in astronomy at the Naval 
Research Laboratory is funded by the Office of Naval Research.  D. J. P. also
acknowledges generous support from the ATNF via a Bolton Fellowship and
from NSF MPS Distinguished International Research Fellowship grant AST0104439.  

We have made extensive use of the HyperLeda database (http://leda.univ-lyon1.fr) and the NASA/IPAC
Extragalactic Database (NED; http://nedwwww.ipac.caltech.edu), which is operated by the Jet Propulsion
Laboratory, California Institute of Technology, under contract with
the National Aeronautics and Space Administration.  We acknowledge the use of 
NASA's {\it SkyView} facility (http://skyview.gsfc.nasa.gov) located at NASA 
Goddard Space Flight Center.  The Digitized Sky
Surveys were produced at the Space Telescope Science Institute under
government grant NAG~W-2166. The images of these surveys are based on
photographic data obtained using the Oschin Schmidt Telescope on
Palomar Mountain and the United Kingdom Schmidt Telescope.

Funding for the Sloan Digital Sky Survey (SDSS) and SDSS-II has been provided by the Alfred P. Sloan Foundation, the Participating Institutions, the National Science Foundation, the U.S. Department of Energy, the National Aeronautics and Space Administration, the Japanese Monbukagakusho, the Max Planck Society, and the Higher Education Funding Council for England. The SDSS Web Site is http://www.sdss.org/.
The SDSS is managed by the Astrophysical Research Consortium for the Participating Institutions. The Participating Institutions are the American Museum of Natural History, Astrophysical Institute Potsdam, University of Basel, University of Cambridge, Case Western Reserve University, University of Chicago, Drexel University, Fermilab, the Institute for Advanced Study, the Japan Participation Group, Johns Hopkins University, the Joint Institute for Nuclear Astrophysics, the Kavli Institute for Particle Astrophysics and Cosmology, the Korean Scientist Group, the Chinese Academy of Sciences (LAMOST), Los Alamos National Laboratory, the Max-Planck-Institute for Astronomy (MPIA), the Max-Planck-Institute for Astrophysics (MPA), New Mexico State University, Ohio State University, University of Pittsburgh, University of Portsmouth, Princeton University, the United States Naval Observatory, and the University of Washington. 

{\it Facilities:} \facility{VLA}, \facility{OVRO--MMA}.

%% Bibliography
\bibliography{ms_referee_revised_v4.bbl}

\clearpage

%%% TABLES 

%
% VLA and OVRO Observing info --> use deluxetable so can rotate
%
\begin{deluxetable}{cccccccc}
\tabletypesize{\scriptsize}
\rotate
\tablecaption{NRAO--VLA and OVRO--MMA Observations
\label{t:VLAOVRO}}
\tablewidth{0pt}
\tablehead{
\colhead{Source} 	&
\colhead{Date}		&
\colhead{Configuration} &
\colhead{Flux}		&
\colhead{Velocity}	&
\colhead{Total}		&
\colhead{Channel Size}	&
\colhead{Time}		
\\	
			&
			&
			&
\colhead{Calibrators}	&
\colhead{(km s$^{-1}$)}	&
\colhead{Channels}	&
\colhead{(km s$^{-1}$)}	&
\colhead{(hours)}	
}
\startdata
{\bf{\emph{NRAO--VLA}}}	& & & & & & & \\
Mrk 297	&	1992 May	&	C	&	3C286	&	4750	&	63	&	10	&	6	\\
Mrk 314	&	1985 April	&	B	&	3C48	&	2081	&	31	&	10	&	5	\\
Mrk 325	&	2002 November	&	C	&	3C48	&	3412	&	63	&	10	&	1	\\
SDSSJ0834+0139	&	2003 October	&	B	&	3C286	&	4296	&	63	&	10	&	12	\\
& & & & & & & \\
{\bf{\emph{OVRO--MMA}}}	& & & & & & & \\
Mrk 297	&	1993 April$-$June	&	4BG	&	3C273, 3C454.3, Neptune, Uranus			&	4718\tablenotemark{1}	&	124	&	5	&	17	\\
Mrk 325	&	2003 March      &	L	&	3C454.3			&	3375	&	120	&	5	&	8.5	\\
	&	2003 May	&	C	&	3C84, 3C454.3		&	3375	&	120	&	5	&	33.3	\\
\enddata
\tablenotetext{1}{~Local standard of rest.}
\end{deluxetable}

\clearpage

%
% Stats from finished images/MOM0 maps
%
\begin{deluxetable}{ccccc}
\tabletypesize{\scriptsize}
\tablecaption{Information on Total Emission Maps
\label{t:VLAOVROmapstats}}
\tablewidth{0pt}
\tablehead{
\colhead{Source}			&
\colhead{Beam Size}		&
\colhead{Emission Velocity Range} 		&
\colhead{rms}		&
\colhead{H~{\tiny{I}} Mass Sensitivity} 			\\
			&
\colhead{(arc sec)}		&
\colhead{(km s$^{-1}$)}		&
\colhead{(Jy beam$^{-1}$ m s$^{-1}$)}		&
\colhead{(M$_\odot$ beam$^{-1}$)}\\
}
\startdata
{\bf{\emph{{H~{\tiny{I}}}}}}&&&&\\
Mrk 297		&	14 $\times$ 13	&	4510 $-$ 4980	&	15	&	1.6 $\times$ 10$^7$ \\
Mrk 314		&	14 $\times$ 11	&	1981 $-$ 2191	&	23	& 5.0 $\times$ 10$^6$	 \\
Mrk 325		&	20 $\times$ 19	&	3292 $-$ 3592 	&	1.0	& 5.7 $\times$ 10$^5$	 \\
SDSSJ0834+0139	&	4.8 $\times$ 4.3	& 4063 $-$ 4413		& 	1.9 &	1.6 $\times$ 10$^6$ \\
&&&&\\
{\bf{\emph{CO(J=1$-$0)}}}&&&&\\
Mrk 297	&	4.6 $\times$ 3.7	&	4693 $-$ 4753	&	57	& 	\\
Mrk 325	&	9.4 $\times$ 6.6	&	3382 $-$ 3432	&	36	& 	\\	
\enddata
\end{deluxetable}

\clearpage

\clearpage
% Stats from ISPEC
%
\begin{deluxetable}{cccccc}
\tabletypesize{\scriptsize}
\tablecaption{Comparison of VLA and GBT \hi\ Integrated Intensities and Masses
\label{t:VLAMHI}}
\tablewidth{0pt}
\tablehead{
\colhead{Source}			&
\colhead{VLA $\int$S dv}		&
\colhead{GBT $\int$S dv}	&
\colhead{VLA M$_{HI}$}		&
\colhead{GBT M$_{HI}$}		&
\colhead{$\tabfrac{VLA~M_{HI}}{GBT~M_{HI}}$}\\
			&
\colhead{(Jy km s$^{-1}$)}	&
\colhead{(Jy km s$^{-1}$)}	&
\colhead{(10$^9$ M$_\odot$)}		&
\colhead{(10$^9$ M$_\odot$)}		&
		\\
}
\startdata
Mrk 297	&	5.8	$\pm$	0.2	&	6.5	$\pm$	0.08	&	6.3	$\pm$	0.2	&	7.0	$\pm$	0.1	&	0.90 $\pm$ 0.04\\
Mrk 314	&	11	$\pm$	0.9	&	12	$\pm$	0.1	&	2.0	$\pm$	0.2	&	2.5	$\pm$	0.03	&	0.80 $\pm$ 0.08	\\
Mrk 325	&	10	$\pm$	0.7	&	11	$\pm$	0.2	&	5.7	$\pm$	0.4	&	6.3	$\pm$	0.1	&	0.90 $\pm$ 0.07	\\
Mrk 538	&				&	20	$\pm$	0.2	&	1.7\tablenotemark{1}			&	7.6	$\pm$	0.1	&	0.22 	\\
SDSSJ0834+0139	&	1.6	$\pm$	0.2	&	6.9	$\pm$	0.2	&	1.4	$\pm$	0.2	&	5.9	$\pm$	0.2	&	0.24 $\pm$ 0.03	\\
\enddata
\tablenotetext{1}{From \cite{1997ApJ...483..754S}}
\end{deluxetable}

\clearpage

%
% W20 and Mdyn (VLA and GBT comparison)
%
\begin{deluxetable}{cccccc}
\tabletypesize{\scriptsize}
\tablecaption{Comparison of VLA and GBT \hi\ Line Widths and Dynamical Masses
\label{t:VLAW20}}
\tablewidth{0pt}
\tablehead{
\colhead{Source}				&
\colhead{VLA W$_{R}^i$}			&
\colhead{VLA M$_{DYN}$\tablenotemark{1}}	&
\colhead{GBT W$_{R}^i$}			&
\colhead{GBT M$_{DYN}$\tablenotemark{1}} 	&
\colhead{$\tabfrac{VLA M_{DYN}}{GBT M_{DYN}}$} \\
				&
\colhead{(km s$^{-1}$)}			&
\colhead{(10$^{10}$ M$_\odot$)}		&
\colhead{(km s$^{-1}$)}			&
\colhead{(10$^{10}$ M$_\odot$)}		&
\\
}
\startdata
Mrk297	&	595			&	17			&	486			&	11			&	1.5	\\
Mrk 314	&	158			&	0.54			&	170			&	0.63			&	0.86	\\
Mrk 325	&	207			&	2.4			&	242			&	3.3			&	0.73	\\
Mrk 538	&	251\tablenotemark{2}			&	4.1			&	264			&	4.5			&	0.91	\\
SDSSJ0834+0139	&	250			&	2.4			&	293			&	3.3			&	0.73	\\
\enddata
\tablenotetext{1}{~Within R$_{25}$}
\tablenotetext{2}{~Estimated from FWHM published in \cite{1997ApJ...483..754S}.}
\end{deluxetable}

\clearpage

%%% FIGURES

%%% START -- 2x2 SUBFIGURE TEMPLATE %%%
%\begin{figure}[hbtp]
%\vspace{-0.5in}
%\centering
%\vspace{.3in}
%\subfigure[
%]
%{\label{f:name1}
%\includegaphics [height = 3in] {name1.eps}}
%\hspace{.3in}
%\subfigure[
%]
%{\label{f:name2}
%\includegraphics [height = 3in] {name2.eps}}
%\hspace{.3in}
%\subfigure[
%]
%{\label{f:name3}
%\includegraphics [height = 3in] {name3.eps}}
%\hspace{.3in}
%\subfigure[
%]
%{\label{f:name4}
%\includegraphics [height=3in] {name4.eps}}
%\vspace{.3in}
%\caption{}
%\label{f:name}
%\end{figure}
%%%END -- 2x2 SUBFIGURE TEMPLATE%%%

%%%BEGIN PANEL 1 -- MRK 297%%%
\begin{figure}[hbtp]
\vspace{-0.5in}
\begin{center}
\vspace{.3in}
{\figurenum{1a}
\label{f:MRK297.HI.MOM0}
\includegraphics [height = 3in] {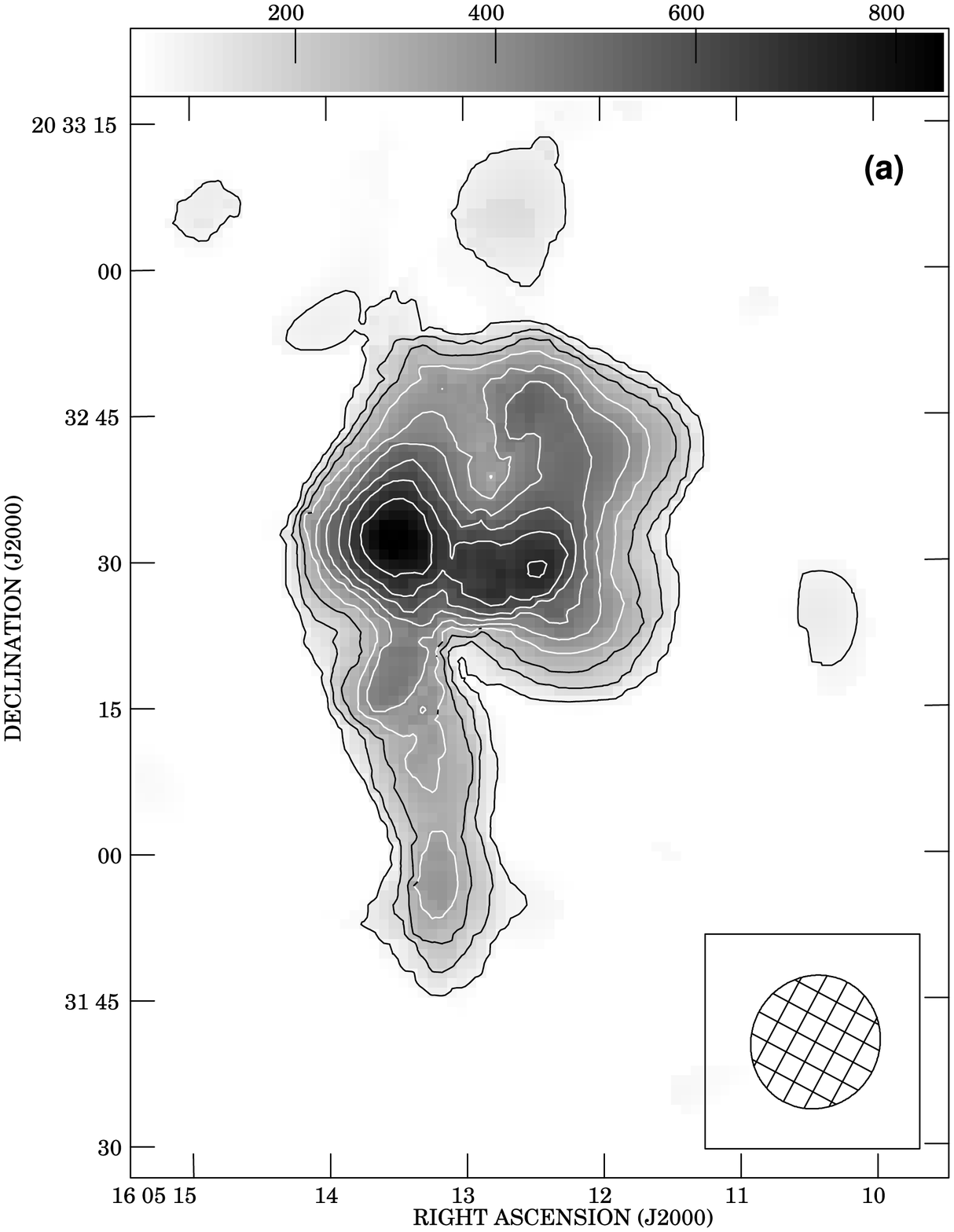}}
\hspace{.3in}
{\figurenum{1b} \label{f:MRK297.HI.SDSS}
\includegraphics [height = 3in] {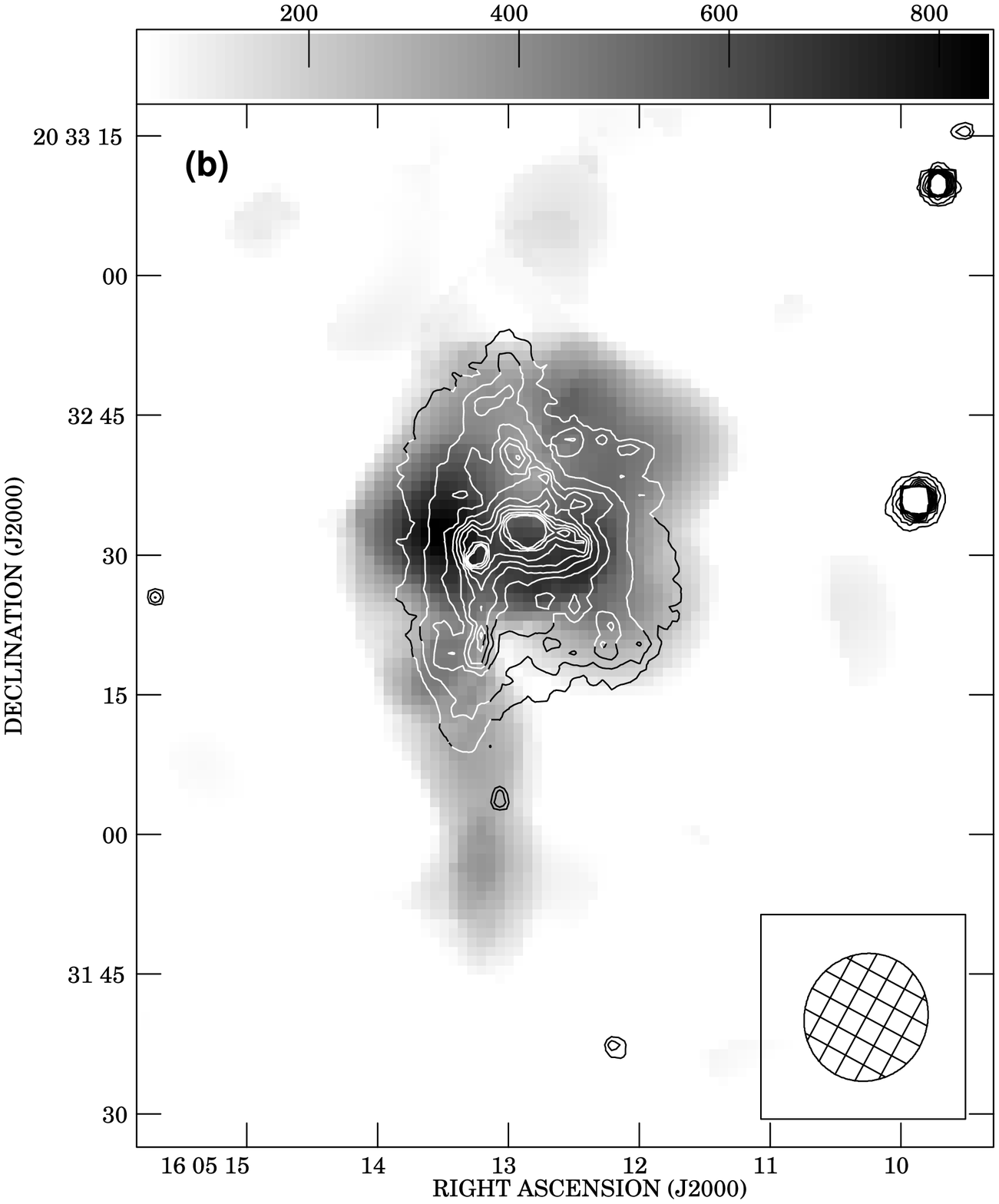}}
\hspace{.3in}
{\figurenum{1c} \label{f:MRK297.HI.MOM1}
\includegraphics [height = 3in] {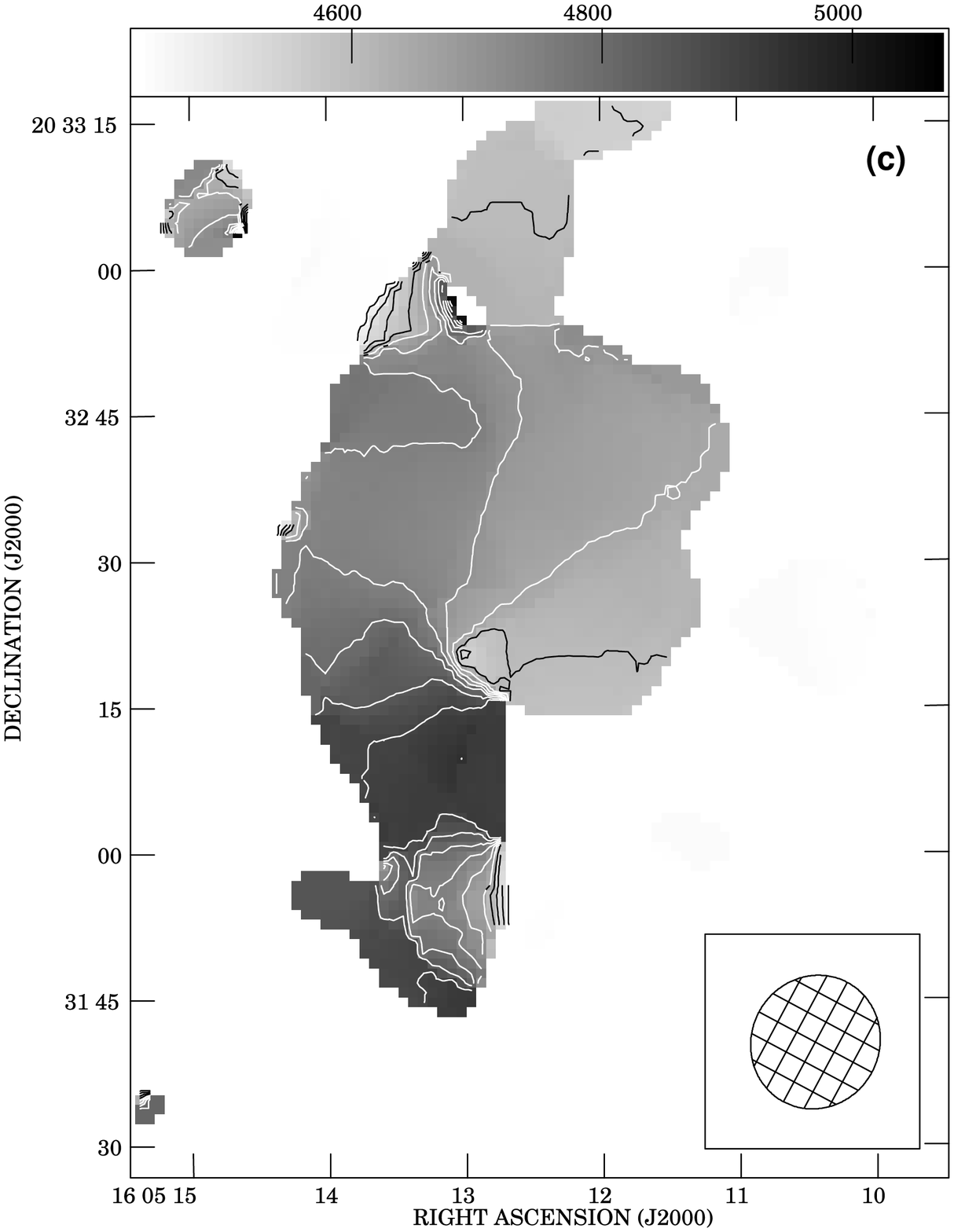}}
\hspace{.3in}
{\figurenum{1d} \label{f:MRK297.HIandCO}
\includegraphics [height=3in] {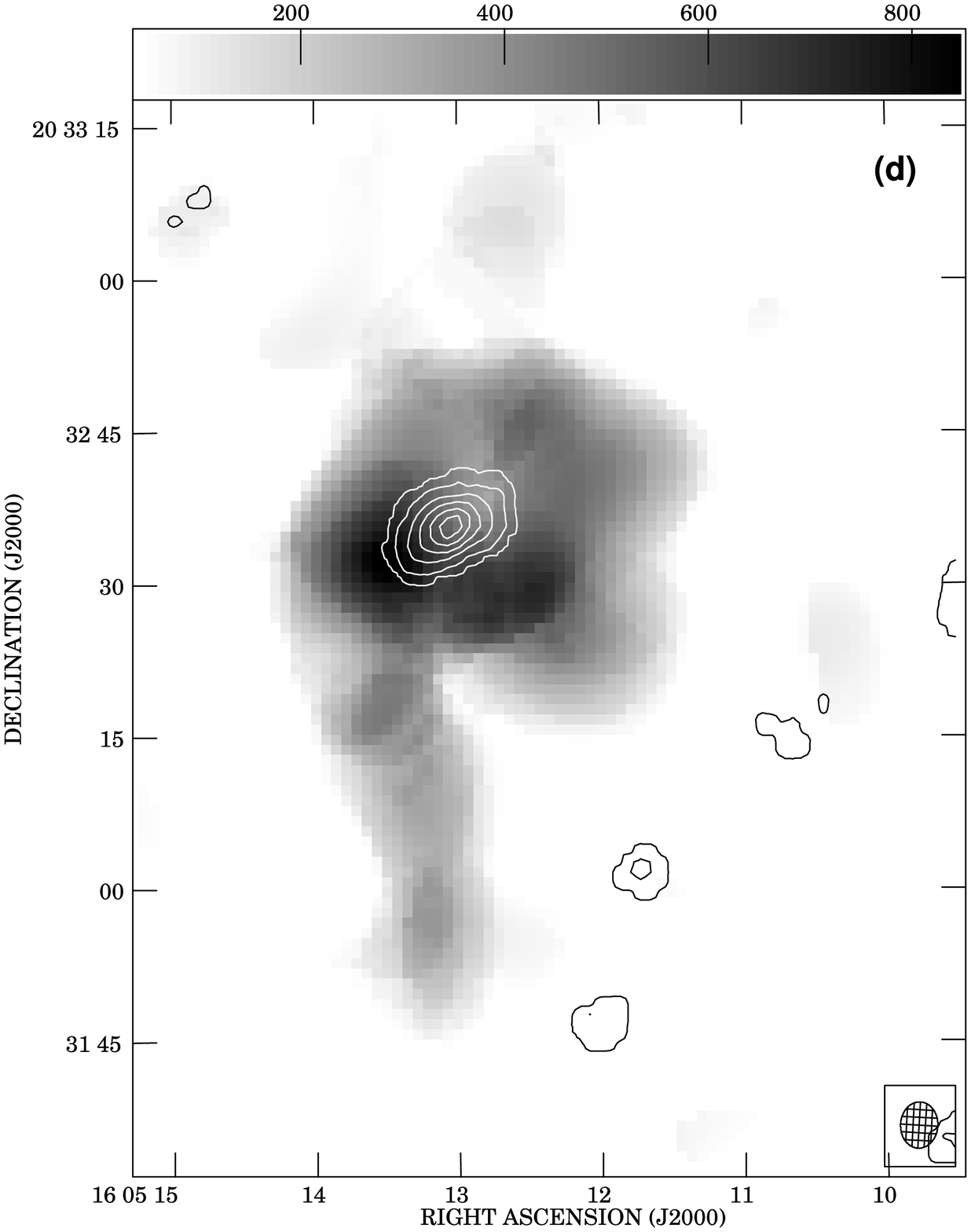}}
\end{center}
\vspace{.1in}
\setcounter{figure}{1} \figurenum{1}
\caption{{\emph{Mrk~297}} 
(a) \hifig\ intensity map.
The first contour and
the contour intervals are at N$_{HI}$~=~5~$\times$~10$^{20}$~cm$^{-2}$.
The bar indicates the gray scale range in units of
Jy~beam$^{-1}$~m~s$^{-1}$; only
\hifig\ intensities above the 3~$\sigma$~level,
45~Jy~beam$^{-1}$~m~s$^{-1}$, are plotted.
The beam is shown in the lower right corner.
(b) Sloan Digital Sky Survey optical image (contours) overlaid on
a gray scale map of the \hifig\ intensity.
The bar indicates the gray scale range
in units of Jy~beam$^{-1}$~m~s$^{-1}$; only
\hifig\ intensities above the 3~$\sigma$~level,
45~Jy~beam$^{-1}$~m~s$^{-1}$, are plotted.  The beam for the \hifig\ map 
is shown in the lower right corner.  The optical contours are arbitrary.  
(c) The \hifig\ velocity field.  The bar indicates
the gray scale range, 4431~$-$~5069 km s$^{-1}$.  The contours
are at 50~km~s$^{-1}$ intervals.  The beam is
shown in the lower right corner.
(d) \co\ intensity (contours) overlaid on a gray scale map
of the \hifig\ intensity.
The contour levels are
1256~$\times$~0.14, 1, 3, 5, 7, and 9 Jy~beam$^{-1}$~m~s$^{-1}$, where 
the first contour corresponds to 3~$\sigma$.
The bar indicates the gray scale range
in units of Jy~beam$^{-1}$~m~s$^{-1}$; only
intensities above the 3~$\sigma$ level,
171~Jy~beam$^{-1}$~m~s$^{-1}$, are plotted.
The \co\ beam is shown in the lower right corner. 
\label{f:MRK297}}
\end{figure}
%%%END PANEL 1 -- MRK 297%%%

\clearpage

%%%BEGIN PANEL 2 -- MRK 314%%%
\begin{figure}[hbtp]
\vspace{-0.5in}
\centering
\vspace{.3in}
{\figurenum{2a} \label{f:MRK314.HI.MOM0}
\includegraphics [height = 2.5in] {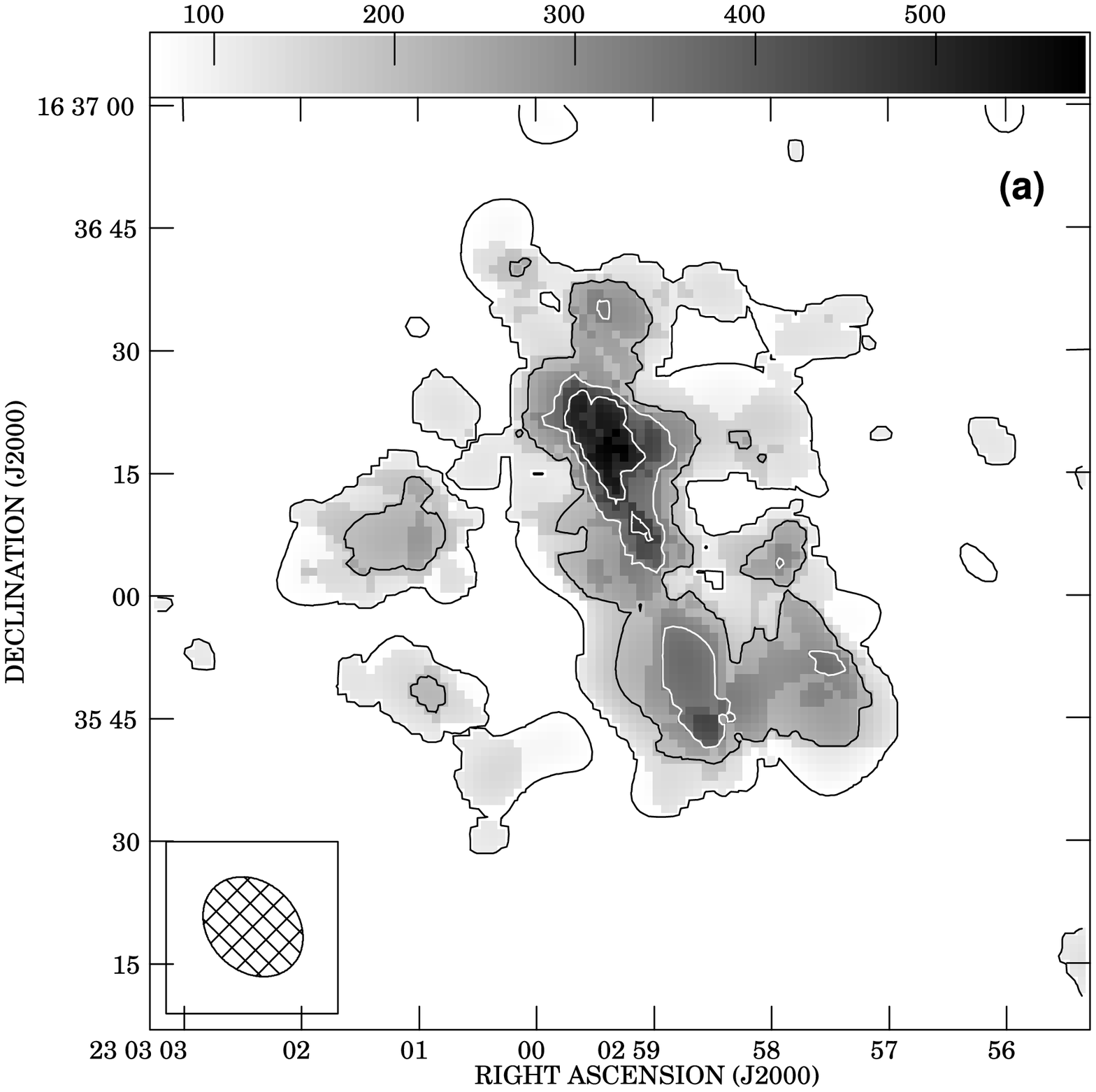}}
\hspace{.3in}
{\figurenum{2b} \label{f:MRK314.HI.DSS}
\includegraphics [height = 2.5in] {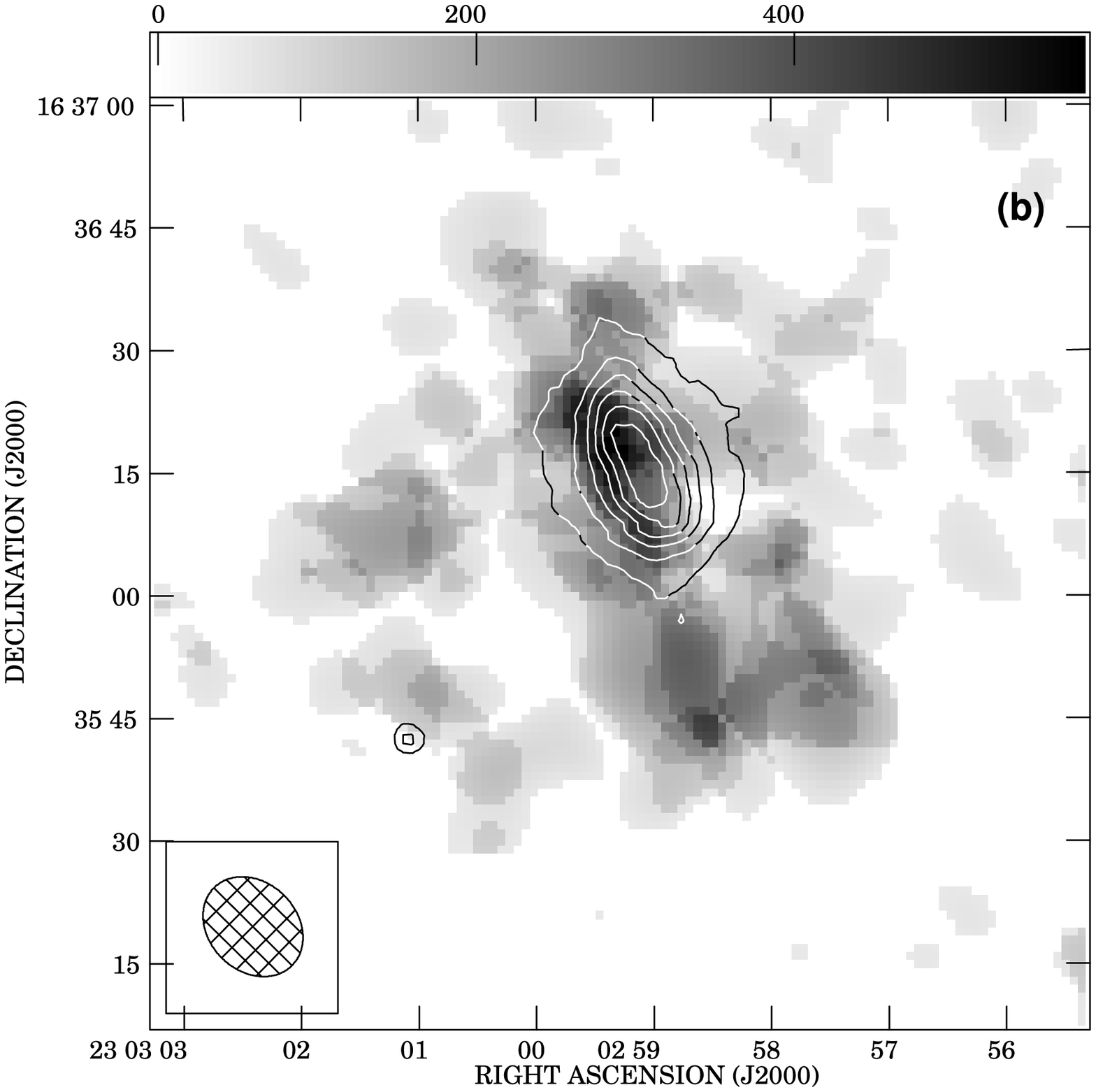}}
\hspace{.3in}
{\figurenum{2c} \label{f:MRK314.HI.MOM1}
\includegraphics [height = 2.5in] {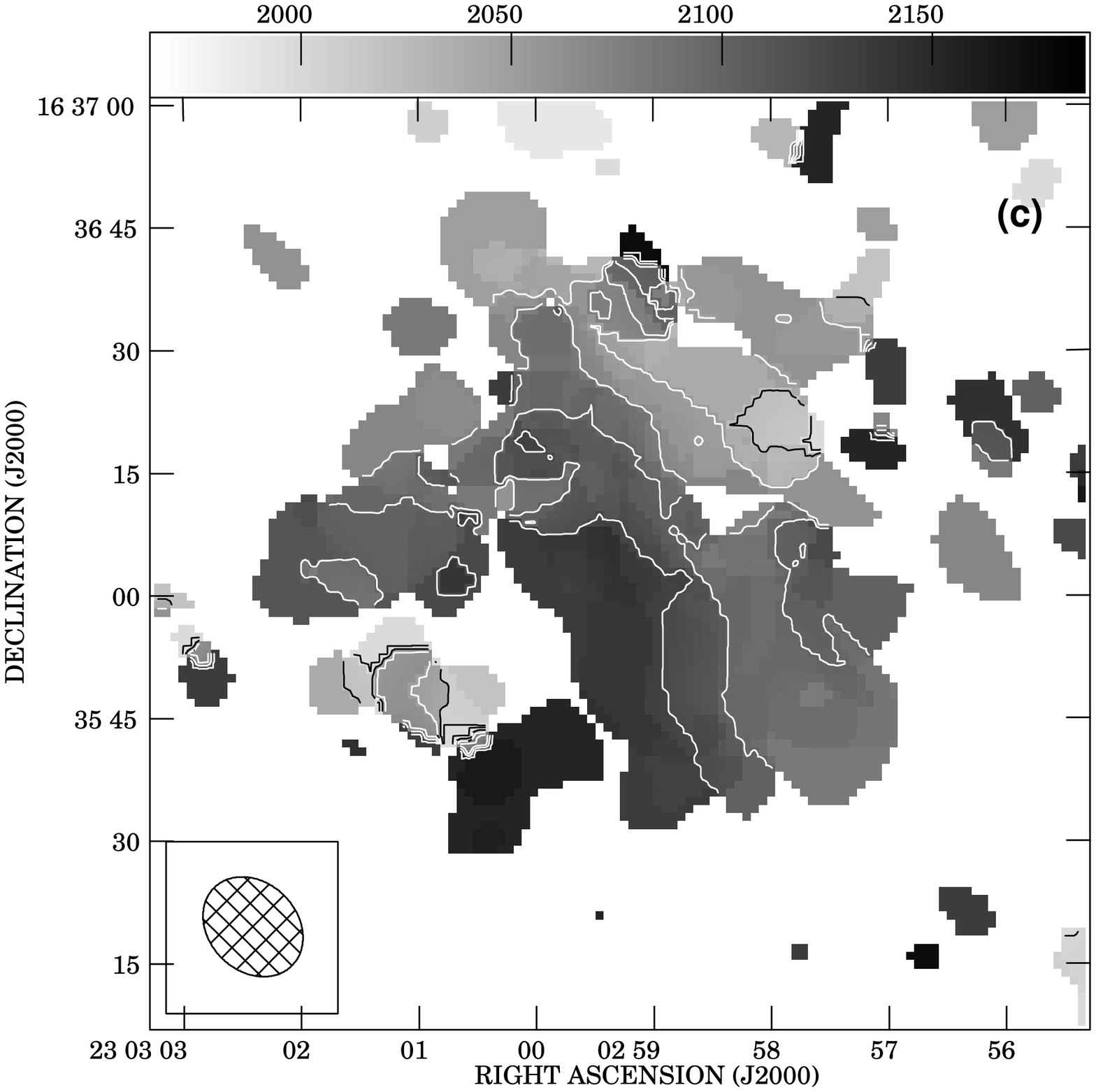}}
\hspace{3in}
\vspace{.3in}
\setcounter{figure}{2} \figurenum{2}
\caption{{\emph{Mrk~314}} 
(a) \hifig\ intensity map. 
The first contour and the contour intervals are at
N$_{HI}$~=~5~$\times$~10$^{20}$~cm$^{-2}$.  
The bar indicates the gray scale range in units of
Jy~beam$^{-1}$~m~s$^{-1}$; only
\hifig\ intensities above the 3~$\sigma$ level, 
69~Jy beam$^{-1}$~m~s$^{-1}$,
are plotted.  
The beam is shown in the lower left corner.
(b) Digitized Sky Survey optical image (contours) overlaid on
a gray scale map of the \hifig\ intensity.
The bar indicates the gray scale range
in units of Jy~beam$^{-1}$~m~s$^{-1}$; only
\hifig\ intensities above the 3~$\sigma$ level, 
69~Jy~beam$^{-1}$~m~s$^{-1}$, are plotted.  The beam 
for the \hifig\ map is shown
in the lower left corner.  The optical contours are arbitrary.  
(c) \hifig\ velocity field.  The bar indicates
the gray scale range, 1966~$-$~2186~km~s$^{-1}$.  
The contour interval is 25~km~s$^{-1}$.
The beam is
shown in the lower left corner.}
\label{f:MRK314}
\end{figure}
%%%END PANEL 2 -- MRK 314%%%

\clearpage

%%%BEGIN PANEL 3 -- MRK 325%%%
\begin{figure}[hbtp]
\vspace{-0.5in}
\centering
\vspace{.3in}
{\figurenum{3a} \label{f:MRK325.HI.MOM0}
\includegraphics [height = 2.5in] {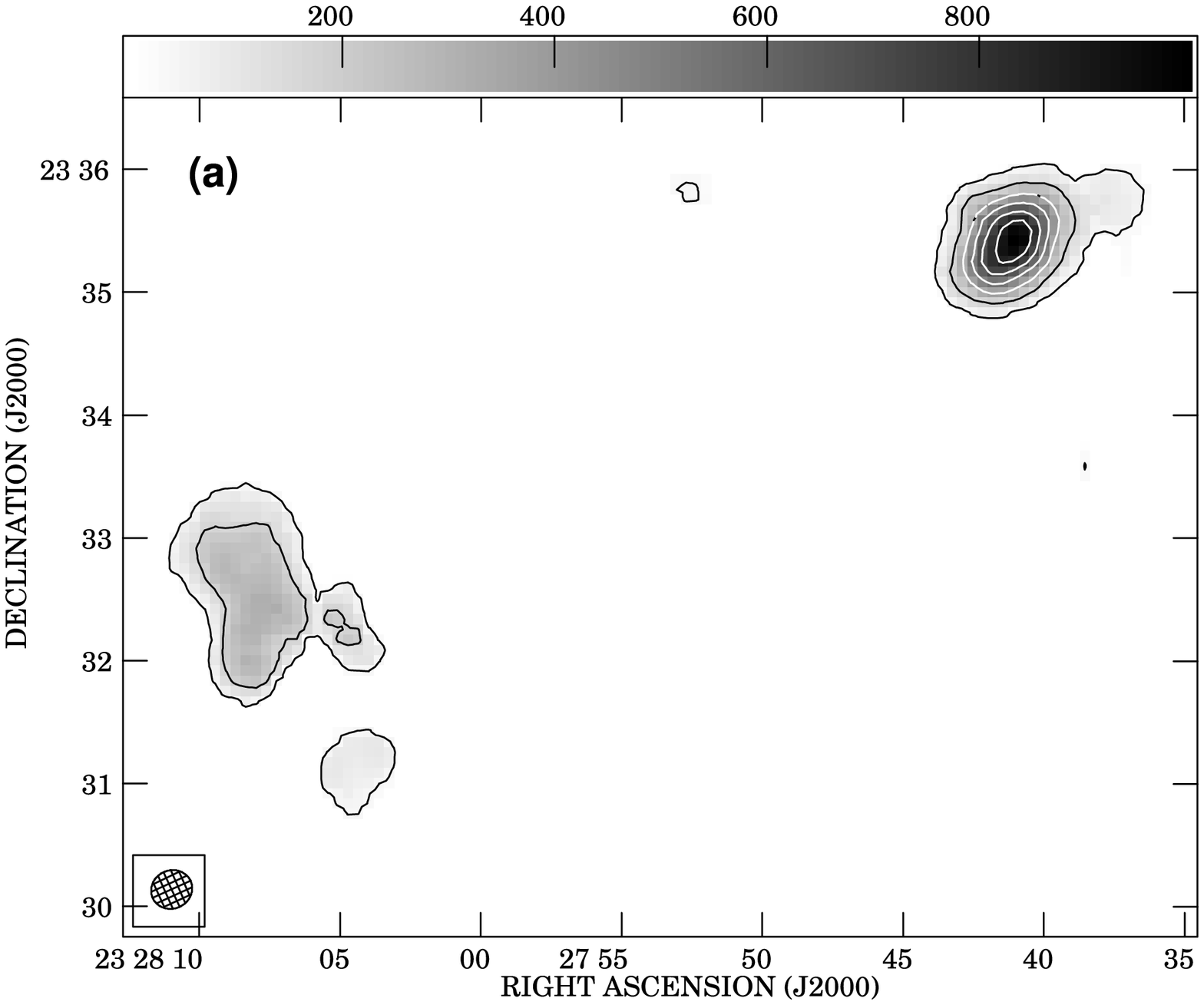}}
\hspace{.3in}
{\figurenum{3b} \label{f:MRK325.HI.DSS}
\includegraphics [height = 2.5in] {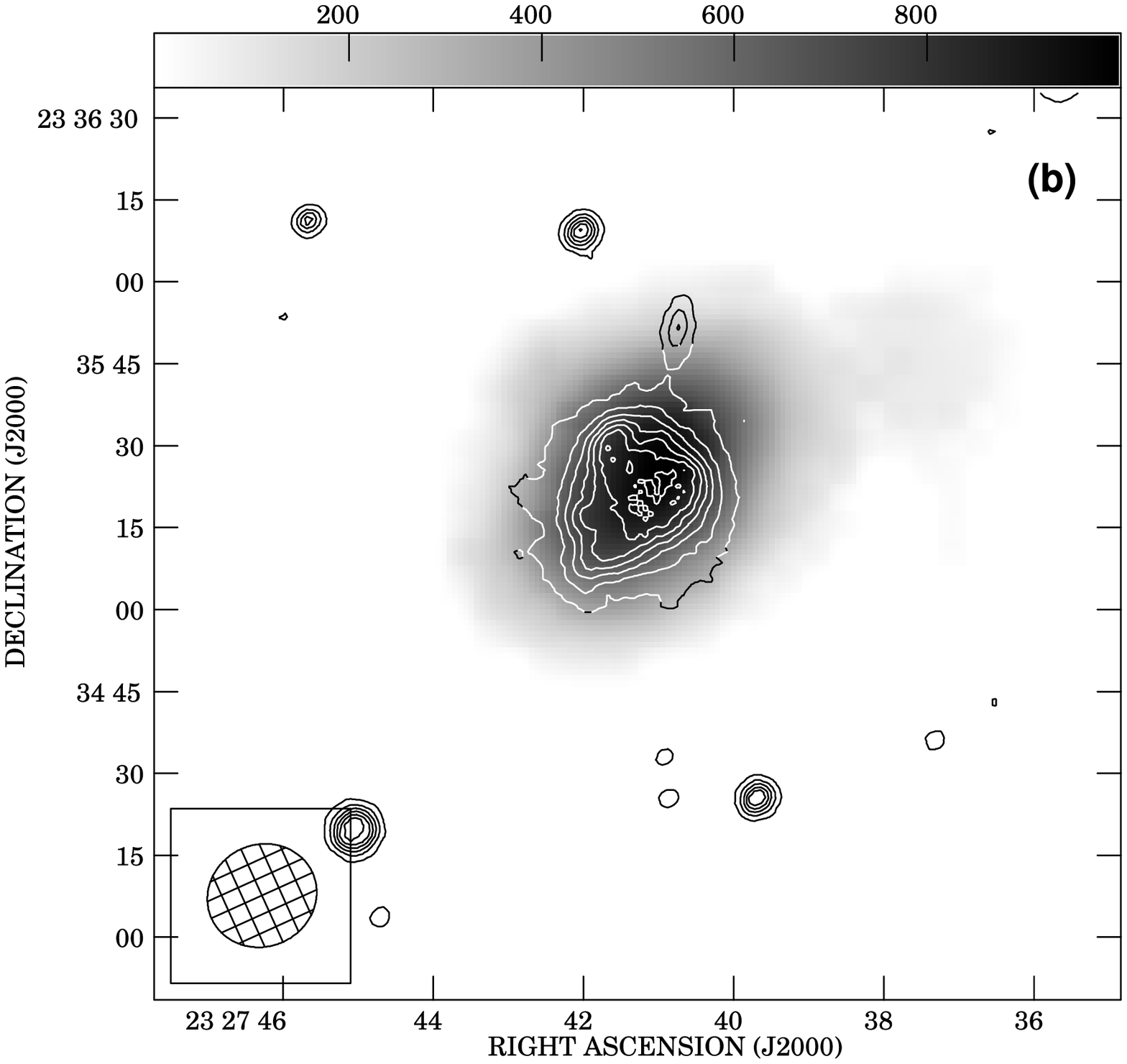}}
\hspace{-1in}
{\figurenum{3c} \label{f:MRK325.HI.MOM1}
\includegraphics [height = 2.5in] {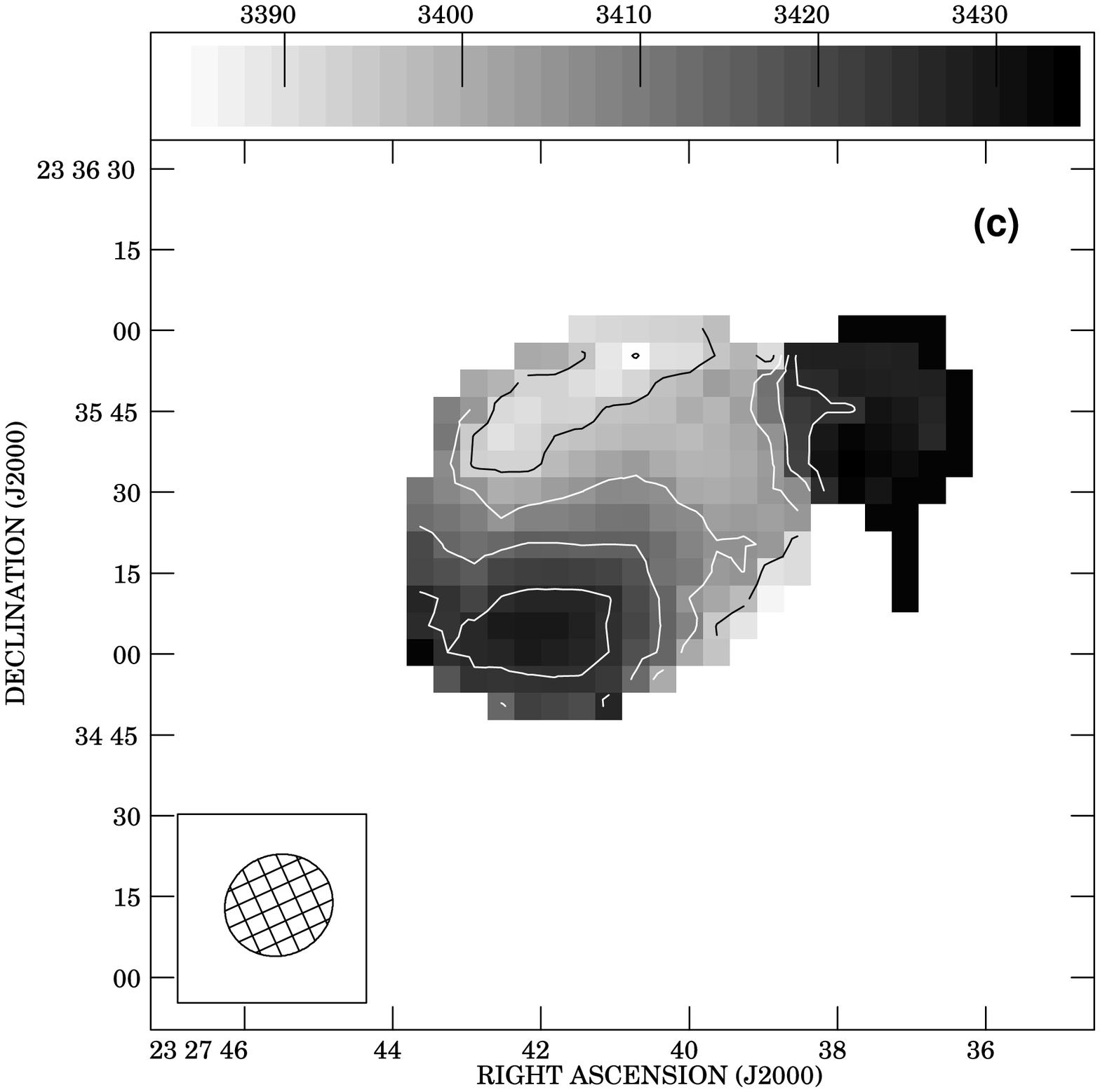}}
\hspace{0.75in}
{\figurenum{3d} \label{f:MRK325.HIandCO}
\includegraphics [height=2.5in] {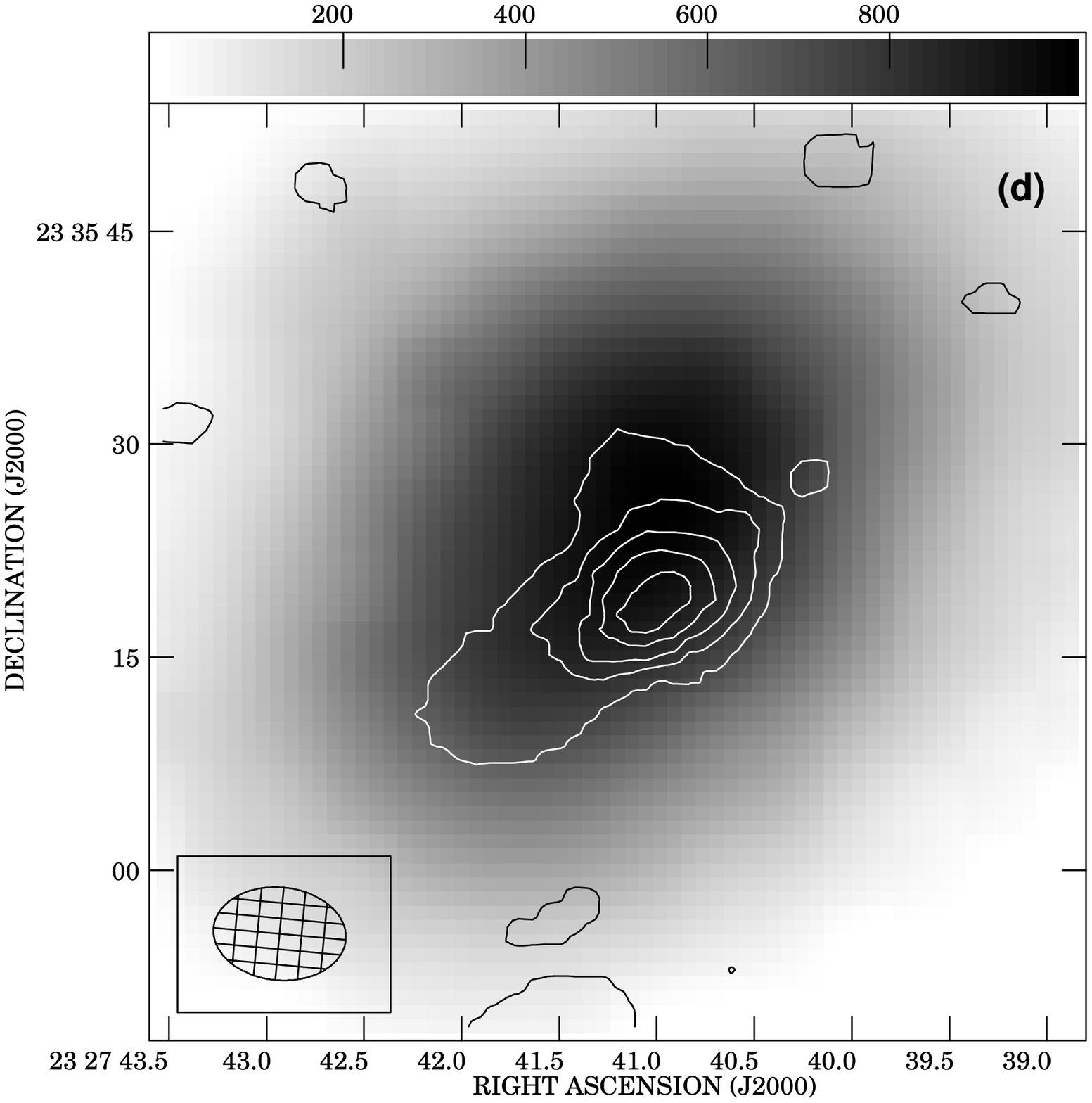}}
\vspace{0.3in}
\setcounter{figure}{3} \figurenum{3}
\caption{{\emph{Mrk~325}} 
(a) \hifig\ intensity map, including both Mrk~325 (right) and Mrk 326.
The first contour and the contour intervals are at 
N$_{HI}$~=~1~$\times$~10$^{20}$~cm$^{-2}$.  
The bar indicates the gray scale range in units of
of Jy~beam$^{-1}$~m~s$^{-1}$; only
\hifig\ intensities above the 3~$\sigma$ level, 
3~Jy~beam$^{-1}$~m~s$^{-1}$,
are plotted.  
The beam is shown in the lower left corner.
(b) Digitized Sky Survey optical image (contours) overlaid on
a gray scale map of the \hifig\ intensity of only Mrk 325.
The bar indicates the gray scale range
in units of Jy~beam$^{-1}$~m~s$^{-1}$; only
\hifig\ intensities above the 3~$\sigma$ level, 
3~Jy~beam$^{-1}$~m~s$^{-1}$, are plotted.  The beam for the \hifig\ map 
is shown in the lower left corner.  The optical contours are arbitrary.  
(c) \hifig\ velocity field of Mrk 325.  The bar 
indicates
the gray scale range, 3385~$-$~3425~km~s$^{-1}$.  
The contour interval is 10~km~s$^{-1}$.
The beam is
shown in the lower left corner.
(d) \co\ intensity (contours) overlaid on a gray scale map
of the \hifig\ intensity for only Mrk 325.
The contour levels
are 418~Jy~beam$^{-1}$~m~s$^{-1}$~$\times$~0.3, 2, 4, 6, and 8.
The first contour corresponds to the 3~$\sigma$ level,
108~Jy beam$^{-1}$~m~s$^{-1}$.
The bar indicates the gray scale range
in units of Jy~beam$^{-1}$~m~s$^{-1}$; only
intensities above the 3~$\sigma$ level, 
3~Jy~beam$^{-1}$~m~s$^{-1}$, are plotted.  
The \hifig\ map was oversampled to match the 
\co\ map. The beam for the \co\ data is shown in the lower left.}
\label{f:MRK325}
\end{figure}
%%%END PANEL 3 -- MRK 325%%%

\clearpage

%%%BEGIN PANEL 4 -- SDSSJ0834+0139%%%
\begin{figure}[hbtp]
\vspace{-0.5in}
\centering
\vspace{.3in}
{\figurenum{4a} \label{f:SL0834.HI.MOM0}
\includegraphics [height = 2.25in] {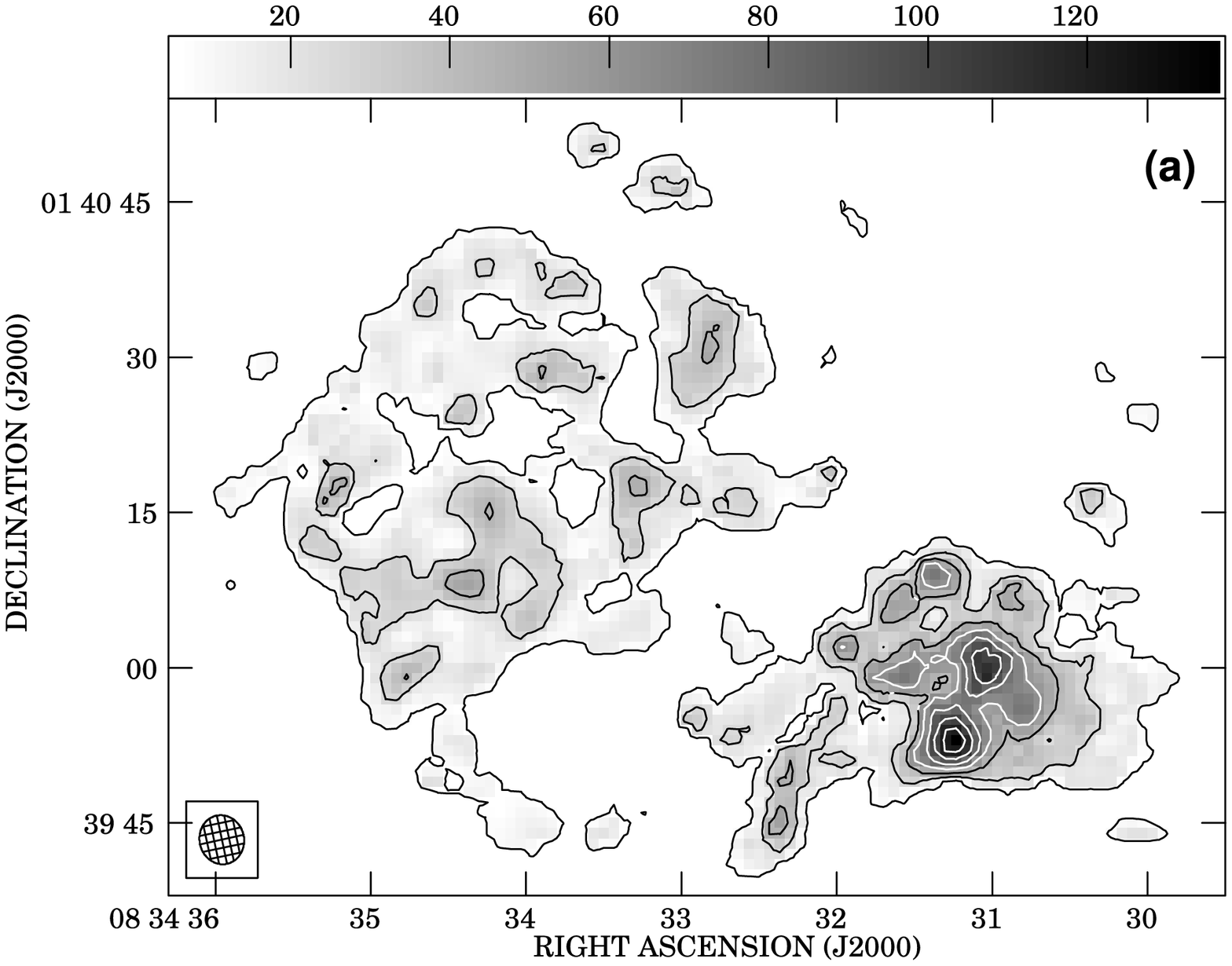}}
\hspace{.3in}
{\figurenum{4b} \label{f:SL0834.HI.SDSS}
\includegraphics [height = 2.25in] {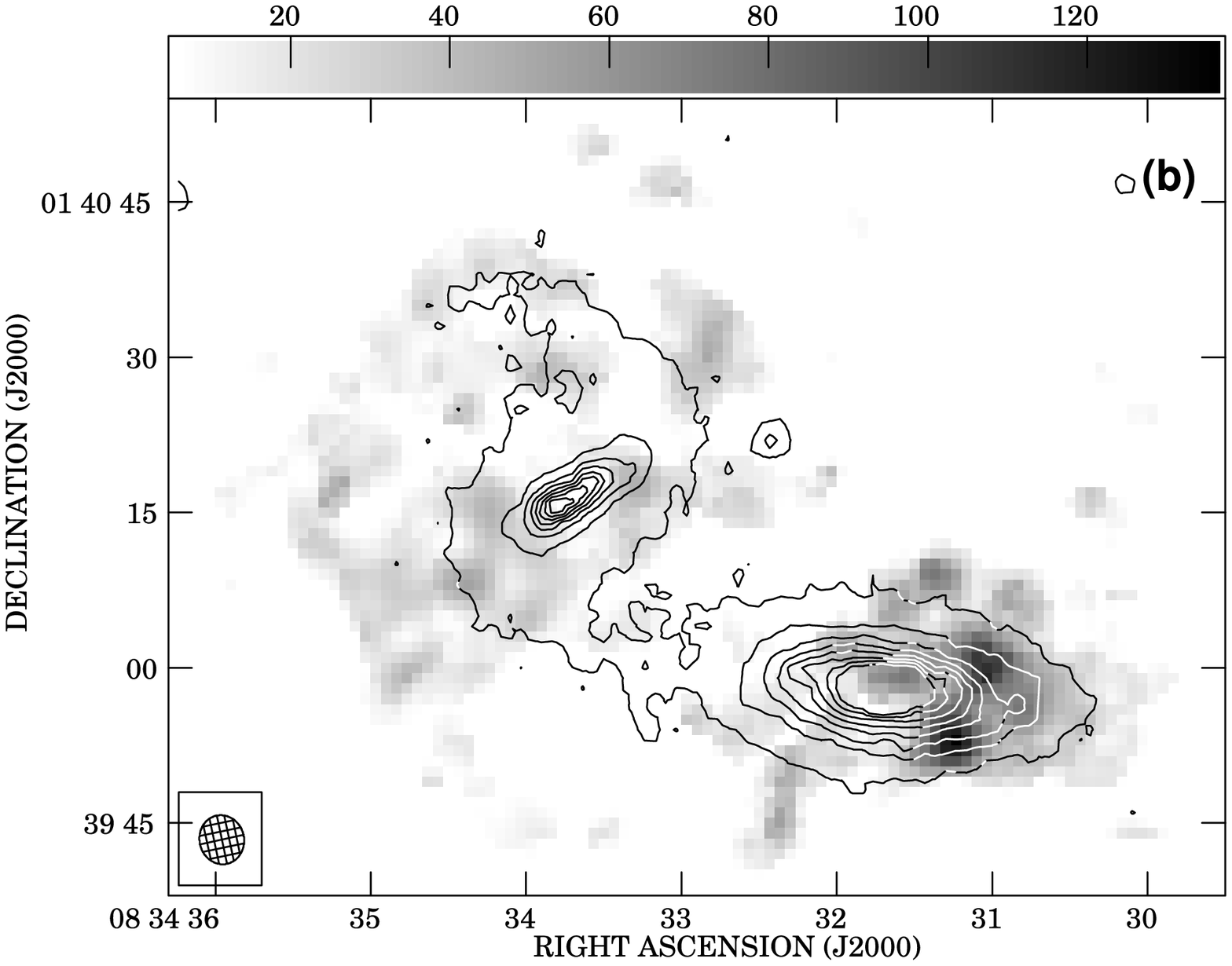}}
\hspace{.3in}
{\figurenum{4c} \label{f:SL0834.HI.MOM1}
\includegraphics [height = 2.25in] {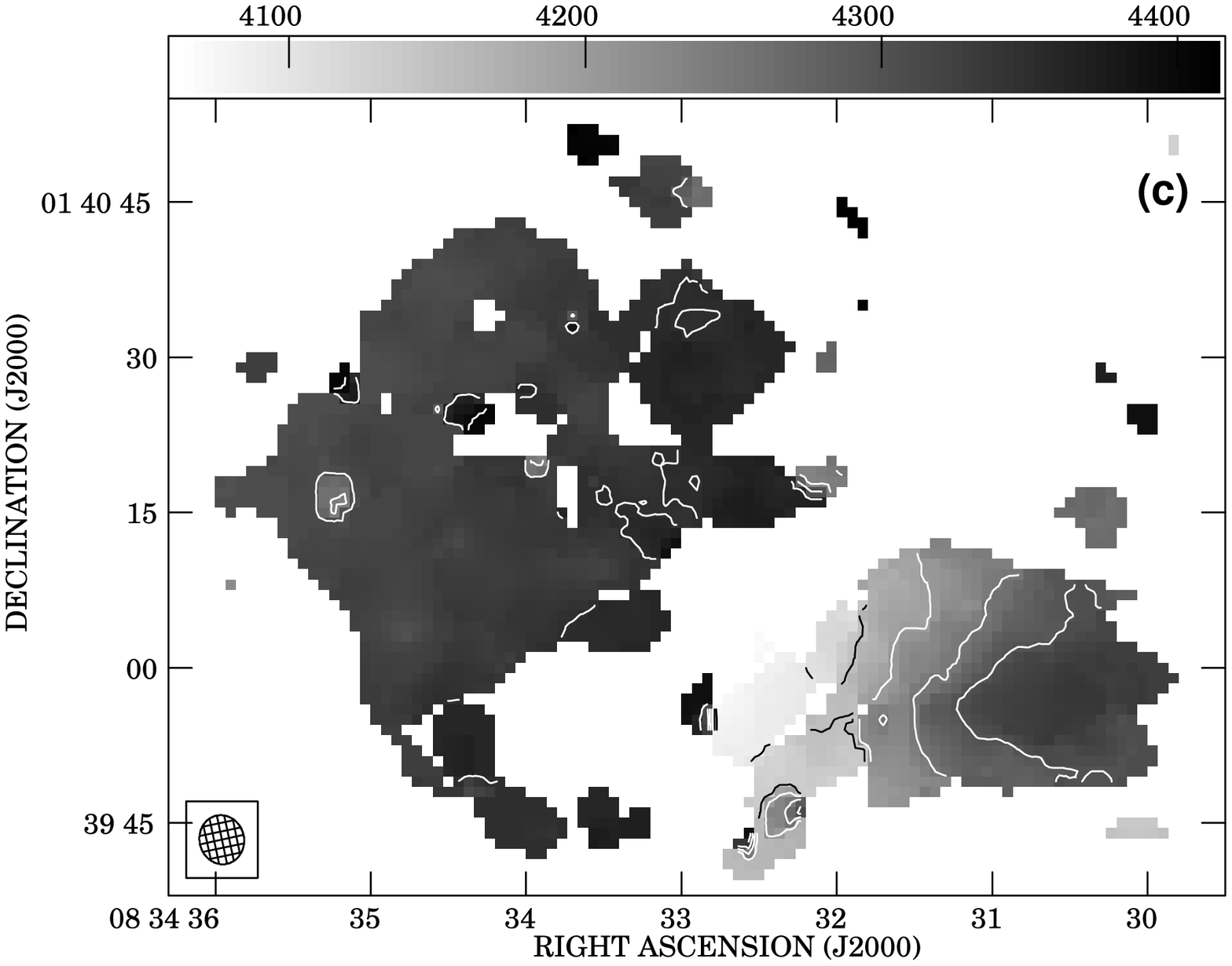}}
\hspace{3.5in}
\setcounter{figure}{4} \figurenum{4}
\caption{{\emph{SDSSJ0834+0139 (right) and companion}} 
(a) \hifig\ intensity map.  The first contour is at 
the 3~$\sigma$ level,  N$_{HI}$~=~3~$\times$~10$^{20}$~cm$^{-2}$.  The 
contour intervals
are at 1~$\times$~10$^{21}$~cm$^{-2}$.
The beam is shown in the lower left corner.
(b) Sloan Digital Sky Survey optical image (contours) overlaid on 
a gray scale map of the \hifig\ intensity.
The bar indicates the gray scale range
in units of Jy~beam$^{-1}$~m~s$^{-1}$; only
\hifig\ intensities above the 3~$\sigma$ level, 
6~Jy~beam$^{-1}$~m~s$^{-1}$, are plotted.  The beam 
for the \hifig\ map
is shown
in the lower left corner.  The optical contours are arbitrary.  
(c) \hifig\ velocity field.  The bar indicates
the gray scale range, 4063~$-$~4413 km~s$^{-1}$.  
The contour interval is 50~km~s$^{-1}$.
The beam is shown in the lower left corner.}
\label{f:SL0834}
\end{figure}
%%%END PANEL 4 -- SDSSJ0834+0139%%%

\clearpage

\begin{figure}
\centerline{
\includegraphics [height=6in] {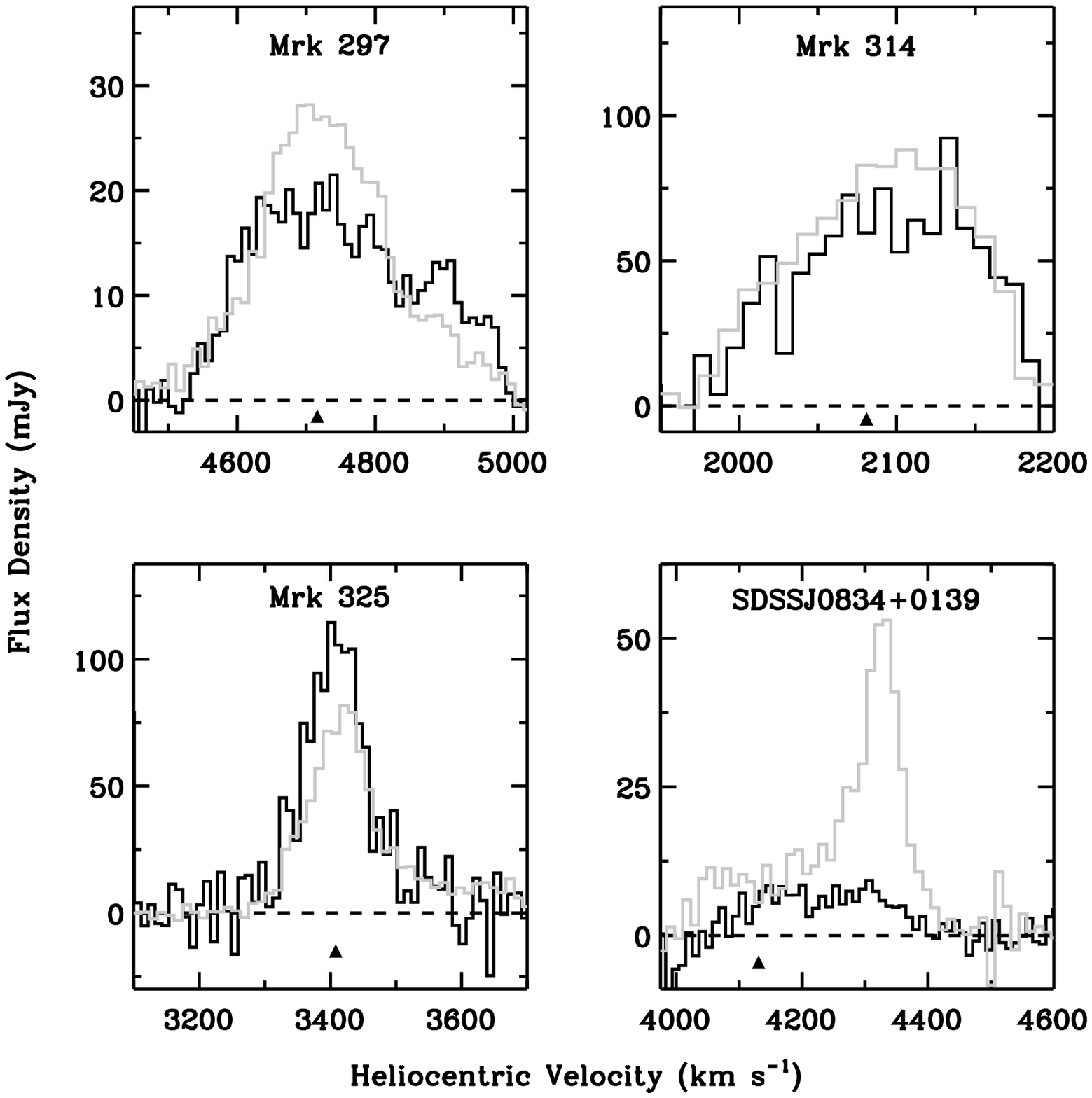}
}
\setcounter{figure}{5} \figurenum{5} 
\caption{\hifig\ spectra of four local LCBGs from both the VLA (black) 
and GBT (gray) observations.  The galaxy name is indicated at the top of
each panel, while the black triangles indicate the recessional velocity of
each galaxy from optical redshifts.  The full VLA velocity range is shown
for each source.
The VLA spectra were created by measuring the \hifig\ flux for each velocity
channel within a box enclosing each galaxy.  The GBT spectra are from Paper I. 
\label{f:VLAHIspectra}}
\end{figure}

\clearpage

\begin{figure}
\centerline{
\includegraphics [height=6in] {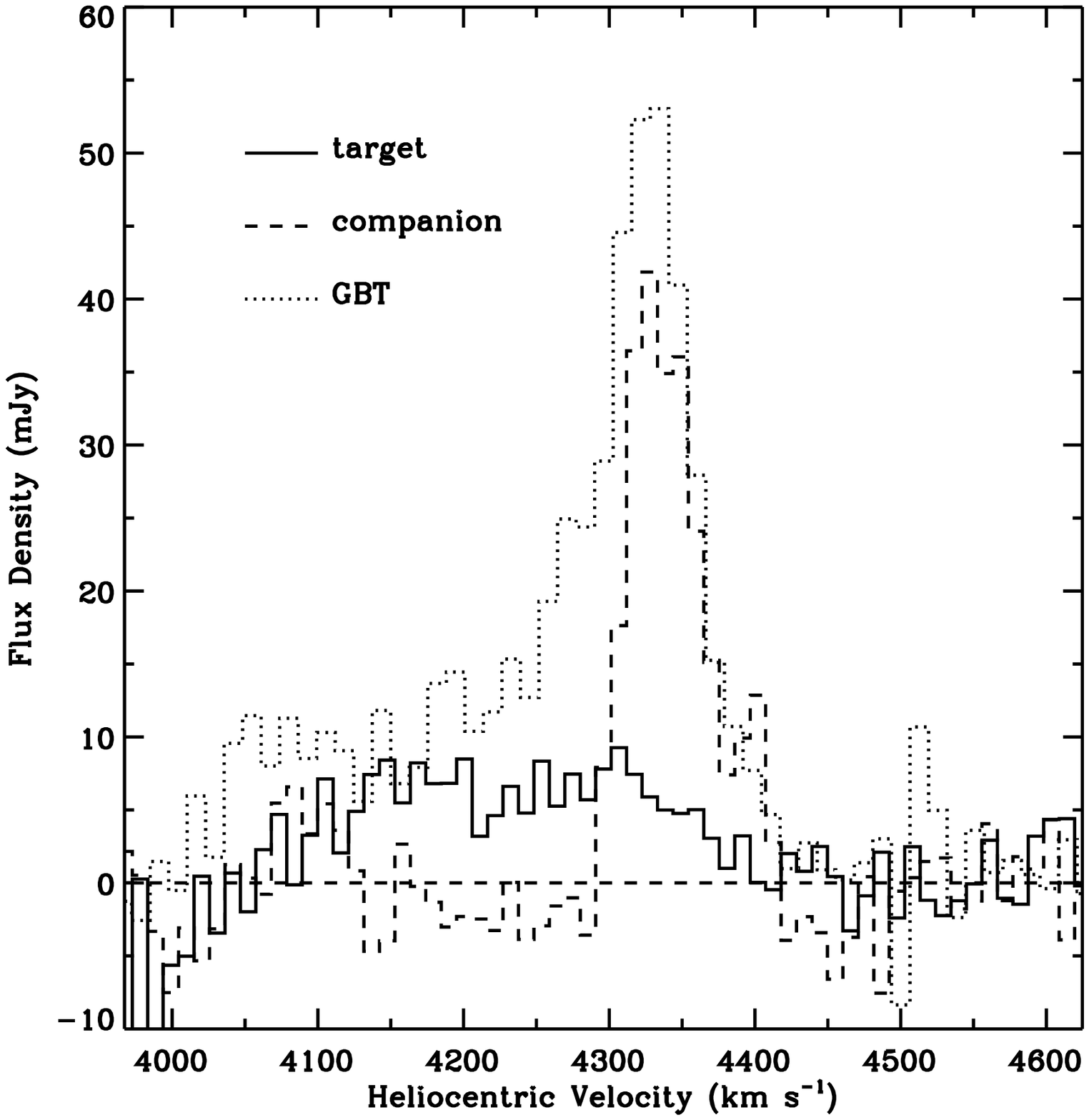}
}
\setcounter{figure}{6} \figurenum{6} 
\caption{Comparison of VLA and GBT \hifig\ spectra of SDSSJ0834+0139 and 
its companion.  The solid line is SDSSJ0834+0139, and the dashed line its 
companion,
from the VLA observations.  The dotted line is the GBT spectrum centered on
SDSSJ0834+0139, but including the companion due to the large beam size of 
the GBT.
\label{f:SL0834decomposed}}
\end{figure}

\clearpage

\begin{figure}
\centerline{
\includegraphics [height=6in] {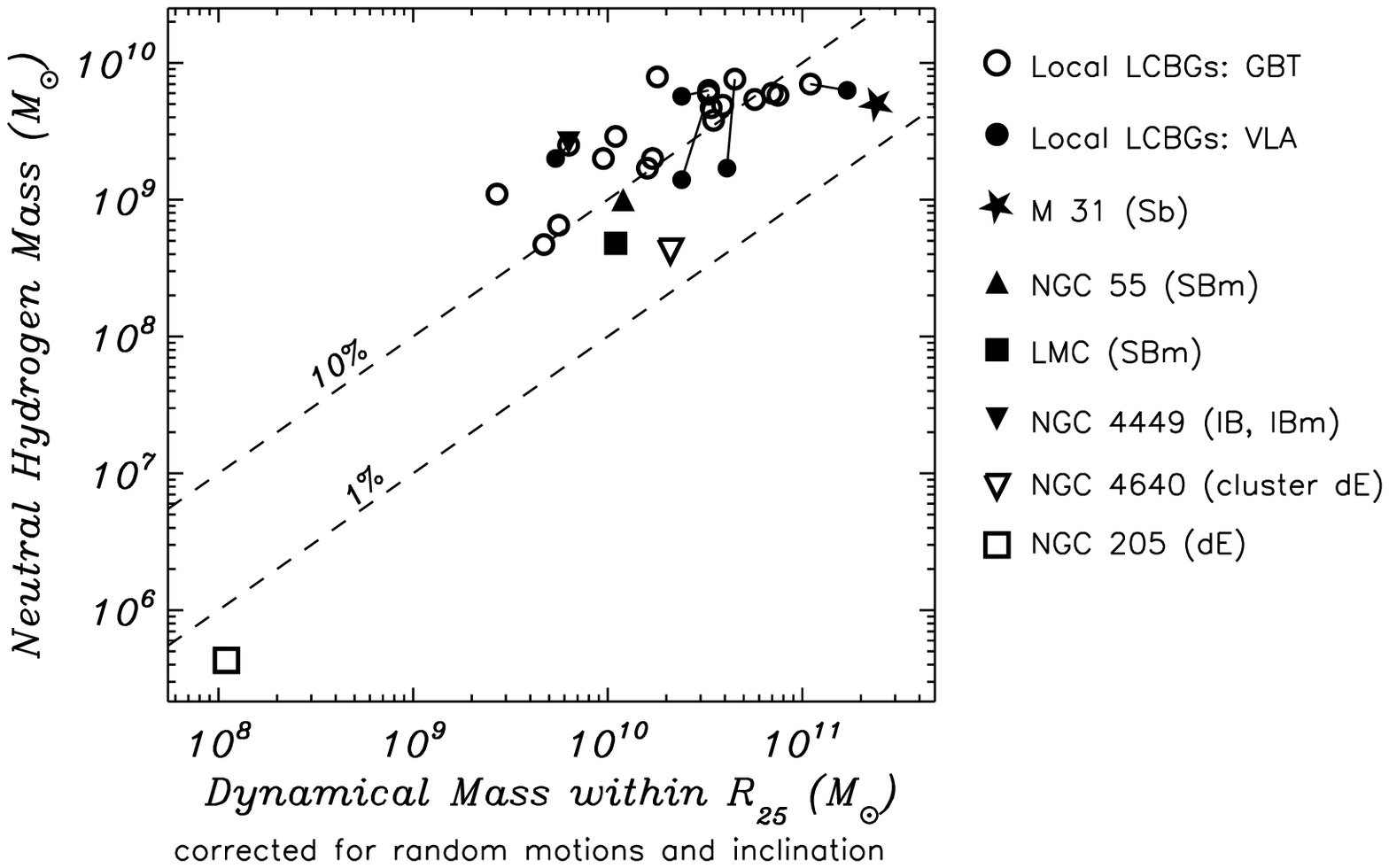}
}
\setcounter{figure}{7} \figurenum{7} 
\caption{Mass of \hi\ versus dynamical mass (within R$_{25}$) for our sample of local LCBGs and other well known galaxies, as indicated on the right.  The galaxy classifications are indicated in parentheses.  A line connects the GBT and VLA derived masses for each LCBG included in this paper. The dynamical masses are calculated from line widths which were corrected both for random motions and inclination. The dashed lines indicate gas mass fractions (M$_{HI}$~M$_{DYN}^{-1}$) of 10 and 1\%.
\label{f:galaxy_comp}}
\end{figure}

\clearpage

\end{document}